\documentclass[superscriptaddress,twocolumn,secnumarabic,
 amssymb,amsmath,nobibnotes,aps,prd,showkeys,showpacs,nofootinbib]{revtex4}

\usepackage{graphicx}
\usepackage{epsf}
\usepackage{bm}%
\newcommand{\be}{\begin{equation}}
\newcommand{\ee}{\end{equation}}

\newcommand{\CC}{\Lambda}

\newcommand{\Omm}{\Omega_m}

\newcommand{\OL}{\Omega_{\Lambda}}

\newcommand{\rmo}{\rho_{m0}}
\newcommand{\rmm}{\rho_{m}}

\newcommand{\mincir}{\raise
-3.truept\hbox{\rlap{\hbox{$\sim$}}\raise4.truept\hbox{$<$}\ }}
\newcommand{\magcir}{\raise
-3.truept\hbox{\rlap{\hbox{$\sim$}}\raise4.truept\hbox{$>$}\ }}

\newcommand{\newnewtext}[1]{\text{#1}}
\newcommand{\newnewnewtext}[1]{\text{#1}}

\begin{document}

\title{Hubble expansion \& Structure Formation in Time Varying Vacuum Models}

\author{Spyros Basilakos}
\affiliation{Academy of Athens, Research Center for Astronomy and Applied Mathematics,
 Soranou Efesiou 4, 11527, Athens, Greece}

\author{Manolis Plionis}
\affiliation{Institute of Astronomy \& Astrophysics, National Observatory of Athens,
Thessio 11810, Athens, Greece and
\\Instituto Nacional de Astrof\'isica, \'Optica y Electr\'onica, 72000 Puebla, Mexico}

\author{Joan Sol\`a}

\affiliation{High Energy Physics Group, Dept. Estructura i
Constituents de la Mat\`eria, Universitat de Barcelona, Diagonal
647, 08028 Barcelona, Catalonia, Spain and
\\Institut de Ci\`encies del Cosmos, UB, Barcelona}

\begin{abstract}
We investigate the properties of the FLRW flat cosmological models
in which the vacuum energy density evolves with time, $\Lambda(t)$.
Using different versions of the $\Lambda(t)$ model, namely quantum
field vacuum, power series vacuum and power law vacuum, we find that
the main cosmological functions such as the scale factor of the
universe, the Hubble expansion rate $H$ and the energy densities are
defined analytically. Performing a joint likelihood analysis of the
recent supernovae type Ia data, the Cosmic Microwave Background
(CMB) shift parameter and the Baryonic Acoustic Oscillations (BAOs)
traced by the Sloan Digital Sky Survey (SDSS) galaxies, we put tight
constraints on the main cosmological parameters of the $\Lambda(t)$
scenarios. Furthermore, we study the linear matter fluctuation field
of the above vacuum models. We find that the patterns of the power
series vacuum $\CC=n_1\,H+n_2\,H^2$ predict stronger small scale
dynamics, which implies a faster growth rate of perturbations with
respect to the other two vacuum cases (quantum field and power law),
despite the fact that all the cosmological models share the same
equation of state (EOS) parameter. In the case of the quantum field
vacuum $\CC=n_0+n_2\,H^2$, the corresponding matter fluctuation
field resembles that of the traditional $\Lambda$ cosmology. The
power law vacuum ($\CC\propto a^{-n}$) mimics the classical
quintessence cosmology, the best fit being tilted in the phantom
phase. In this framework, we compare the observed growth rate of
clustering measured from the optical galaxies with those predicted
by the current $\Lambda(t)$ models. Performing a Kolmogorov-Smirnov
(KS) statistical test we show that the cosmological models which
contain a constant vacuum ($\Lambda$CDM), quantum field vacuum and
power law vacuum provide growth rates that match well with the
observed growth rate. However, this is not the case for the power
series vacuum models (in particular, the frequently adduced
$\CC\propto H$ model) in which clusters form at significantly
earlier times ($z\ge 4$) with respect to all other models ($z\sim
2$). Finally, we derived the theoretically predicted dark-matter
halo mass function and the corresponding distribution of
cluster-size halos for all the models studied.
Their expected redshift distribution indicates that it will be difficult to
distinguish the closely resembling models (constant vacuum, quantum
field and power-law vacuum), using realistic future X-ray surveys
of cluster abundances. However, cluster surveys based on the Sunayev-Zeldovich
detection method give some hope to distinguish the closely resembling
models at high redshifts.

\end{abstract}
\pacs{98.80.-k, 95.35.+d, 95.36.+x}
\keywords{Cosmology; dark matter; dark energy}
\maketitle

\section{Introduction}
Over the past decade, studies of the available high quality
cosmological data (supernovae type Ia, CMB, galaxy clustering, etc.)
have converged
towards a cosmic expansion history that involves a spatially flat
geometry and a recent accelerating expansion of the universe (cf.
\cite{Spergel07,essence,komatsu08,Teg04,Kowal08,Hic09} and
references therein). This expansion has been attributed to an energy
component called dark energy (DE) with negative pressure, which
dominates the universe at late times. The simplest type of DE
corresponds to the cosmological constant (see
\cite{Weinberg89,Peebles03,Pad03} for reviews). The so-called
concordance model (or
$\Lambda$CDM model), which contains cold dark matter (DM) to
explain clustering, flat spatial geometry and a cosmological
constant, $\Lambda$, fits accurately the current observational data
and thus it is an excellent candidate to be the model that describes
the observed universe. However, the concordance model suffers from,
among others \cite{Peri08}, two fundamental problems: (a) {\it
The ``old'' cosmological constant problem} (or {\it
fine tuning problem}) i.e., the fact that the observed value of the
vacuum energy density ($\rho_{\Lambda}=\Lambda c^{2}/8\pi G\simeq
10^{-47}\,GeV^4$) is many orders of magnitude below the value found
using quantum field theory (QFT) \cite{Weinberg89}, and (b) {\it the
coincidence problem}\,\cite{coincidence} i.e., the fact
that the matter energy density and the vacuum energy density are of
the same order just prior to the present epoch, despite
the fact that the former is a rapidly decreasing function
of time while the latter is just stationary. The extremal
possibility concerning problem (a) occurs when the Planck
mass~$M_{P}\sim 10^{19}\,\text{GeV}$ is used as the fundamental
scale; then the ratio $M_P^4/\rho_{\Lambda}$ becomes $\sim
10^{123}$. One may think that physics at the Planck scale is not
well under control and that this enormous ratio might be fictitious.
However, let us consider the more modest scale $v=2\,M_W/g\simeq
250\,\text{GeV}$ associated to the electroweak Standard Model (SM)
of Particle Physics (the experimentally most successful QFT known to
date), where $M_W$ and $g$ are the $W^{\pm}$ boson mass and the
$SU(2)$ gauge coupling, respectively. In this case, the ratio of the
predicted vacuum energy versus the measured one is still very large:
it reads $|\langle V\rangle|/\rho_{\Lambda}\gtrsim 10^{55}$, where
$\langle V\rangle= -(1/8)M_H^2\,v^2$ is the vacuum expectation value
of the Higgs potential and $M_H\gtrsim 114.4\,\text{GeV}$ is the
lower bound on the Higgs boson mass in the SM.

Attempts to solve the above problems have been presented in the
literature (see \cite{Peebles03,Pad03,Egan08} and references
therein). Historically, the first attempts to solve the
``old $\CC$ problem'' (the toughest one in the list) were based on
dynamical adjustment mechanisms\,\cite{oldCC} aiming to avoid a mere
fine tuning of the various QFT contributions. Later on it was
realized that a dynamical $\CC$ could also be useful to overpass the
coincidence problem. The idea is to replace the constant vacuum
energy either with a DE that evolves with time
(quintessence and the like\,\cite{Peebles03}), or
alternatively with a time varying vacuum energy density,
$\rho_{\Lambda}(t)$ (cf.\,\cite{Grande06,Bas09b}). In the
quintessence context, one can introduce an
ad hoc tracker scalar field $\phi$\,\cite{Caldwell98} (different
from the usual SM Higgs field) rolling down the potential energy
$V(\phi)$, and therefore it could resemble the DE.
Detailed analysis of these models exist in the literature,
including their confrontation with the
data\,\cite{Jassal,SR,Xin,SVJ} -- see also
\cite{Peebles03,Pad03,Cope06} for extensive reviews. Nevertheless,
the idea of a scalar field rolling down some suitable potential does
not really solve the problem because the initial value of the DE
still needs to be fine tuned.
Moreover, the typical mass of $\phi$ is usually very small.
Indeed, being $\phi$ unrelated to the SM physics, it is usually
thought of as a high energy field $\langle \phi\rangle\simeq M_X$
where $M_X$ is some scale near the Planck mass. If one assumes the
simplest form for its potential, namely
$V(\phi)=m_{\phi}^2\,\phi{^2}/2$, and requires it to reproduce
the present value of the vacuum energy density, $\rho_{\CC}=\langle
V(\phi) \rangle\sim 10^{-11}\,eV^4$, the mass of $\phi$ is predicted
in the ballpark of $m_{\phi}\sim\,H_0\sim 10^{-33}\,eV$. This is an
inconceivably small mass scale in Particle Physics. Thus, the
problem that one is creating along with the introduction of $\phi$
is much bigger than the problem that one is intending to solve, for
one is postulating a mass scale which is $30$ orders of magnitude
below the mass scale $m_{\Lambda}$ associated to the value of the
vacuum energy density $m_{\Lambda}\equiv\rho_{\Lambda}^{1/4}\sim
2.3\times 10^{-3}\,eV$.

Current observations do not rule out the possibility of a
dynamical
DE\,\cite{Spergel07,essence,komatsu08,Teg04,Kowal08,Hic09}.
They indicate that the dark energy EOS parameter
 $w\equiv P_{DE}/\rho_{DE}$ is close to $-1$ to within $\pm
10\%$, if it is assumed to be constant, whilst it is much more poorly
constrained if it varies with time\,\cite{essence}.
Let us note, interestingly enough, that models with running vacuum
energy, may appear in practice with a non-trivial ``effective EOS''
$w=w(t)$\,\cite{SS12}, which can be accommodated by observations.
Indeed, the basic feature in this cosmological ideology is that,
although the EOS parameter of the vacuum energy is strictly equal to
-1, a time evolving $\Lambda$ generally results in an effective EOS,
usually a function $w=w(a)$ of the scale factor that approaches
$w\to -1$ at the present time. This effective EOS may be the result
either of the fact that we are trying to describe the variable
$\Lambda(t)$ model as if it were a quintessence type model with
conserved DE\,\cite{SS12}, or of the existence of other DE entities
mixed up with the variable $\Lambda(t)$, as e.g. in the case of the
$\Lambda$XCDM model\,\cite{Grande06,FSS09a}.

Although the precise functional form of $\Lambda(t)$ is still
missing, an interesting QFT approach within the context of
the renormalization group (RG) was proposed long ago\,
\cite{NelPan82,Toms83}. Later on, the RG-running framework was
further explored in\,\cite{Shap00} from the viewpoint of QFT in
curved space-time by employing the standard perturbative
RG-techniques of Particle Physics (see also \cite{Reuter00} for a
functional RG approach in a non-perturbative context). A more
phenomenological point of view was addressed in
\cite{Oze87}, in which a time varying $\Lambda(t)$ parameter could
be a possible candidate to solve the two fundamental cosmological
puzzles. There is an extensive literature on time varying $\CC$
models (cf.\,\cite{Carvalho92,Free187,Bauer05} and references
therein).

It is worth noticing, that for an important class of $\Lambda(t)$
models there is a coupling between the time-dependent vacuum and
matter. A first work confronting supernovae data with a
(RG-inspired\,\cite{Shap00}) model of this kind was presented
in\,\cite{RGTypeIa} -- see also \cite{meng05}. Using the combination
of the conservation of the total energy with the variation of the
vacuum energy, one can show that these $\Lambda(t)$ models provide
either a particle production process or that the mass of the dark
matter particles increases. Despite the fact that most of the recent
papers on these matters are based on the assumption that the DE
evolves independently of the dark matter, the unknown nature of the
DE and DM implies that at the moment we can not exclude the
possibility of interactions in the dark sector, whether at the level
of a variable $\Lambda(t)$\,\cite{RGTypeIa,meng05} or from coupled
quintessence\,\cite{Zim01}.
Another possibility is that matter is strictly conserved and that
both $\Lambda(t)$ and the gravitational coupling $G(t)$ are running
\,\cite{SSS04,Fossil07}, but this option will not be scrutinized
here.

The scope of the present work is to study the observational
consequences of the overall dynamics for a wide class of
time varying vacuum energy models, in interaction with matter, in the
light of the most recent cosmological data. The structure of the
paper is as follows. The basic theoretical elements of the problem
are presented in section 2, where we introduce [for a spatially flat
Friedmann-Lema\^\i tre-Robertson-Walker (FLRW) geometry] the basic
cosmological equations. In sections 3 and 4 we place constraints on
the main parameters of our vacuum models by performing a joint
likelihood analysis utilizing the so called 'Constitution set' of
the SNIa data \cite{Hic09}, the shift parameter of the Cosmic
Microwave Background (CMB)\,\cite{komatsu08} and the observed
Baryonic Acoustic Oscillations (BAOs; \cite{Eis05}). Also, we
investigate whether the current vacuum cosmological models can yield
a late accelerated phase of the cosmic expansion. In section 5 we
compare the observed linear growth rate of clustering measured from
the SDSS galaxies with those predicted by the $\Lambda(t)$ models,
explored here, while in section 6 we present the corresponding
theoretical predictions regarding the formation of the galaxy
clusters and the evolution of their abundance.
In section 7 we draw our conclusions. Note, that
throughout the paper we use $H_{0}=70.5$ km/sec/Mpc \cite{Freed01,
komatsu08}.

\section{Cosmology with a time dependent vacuum}
In the framework of a spatially flat FLRW geometry the basic
equations which govern the global dynamics of the universe are

\be \rho_{m}+\rho_{\Lambda}=3H^{2} \label{frie1} \ee and \be
\dot{\rho_{m}}+\dot{\rho_{\Lambda}}+ 3H(\rho_{m}+P_{m}+
\rho_{\Lambda}+P_{\Lambda})=0\,,
\label{frie2} \ee

where $\rho_{m}$ is the matter energy density and $\rho_{\Lambda}$
is the vacuum energy density, while $P_{m}$ and $P_{\Lambda}$ are
the corresponding matter and vacuum pressures. Notice, that in order
to simplify our formalism we use geometrical units ($8\pi G=c\equiv
1$) in which $\rho_{\Lambda}=\Lambda$.

In the present work, we would
like to investigate the potential of a time varying
$\Lambda=\Lambda(t)$ parameter to account for the observed
acceleration of the expansion of the Universe. Within this framework
it is interesting to mention that the equation of state takes the
usual form  $P_{\Lambda}(t)=-\rho_{\Lambda}(t)=-\Lambda(t)$, and we
remark that this EOS does not depend on whether $\CC$ is strictly
constant or variable. Also, by introducing in the global dynamics
the idea of the time-dependent vacuum, it is possible to explain the
physical properties of the DE
as well as to ameliorate the status of the fine tuning and
the coincidence problem respectively. In the matter
dominated epoch ($P_{m}=0$), eq.(\ref{frie2}) leads to the
following useful formula:
\begin{equation}
\dot{\rho}_{m}+3H\rho_{m}=-\dot{\Lambda}\,, \label{frie33}
\end{equation}
and considering eq.(\ref{frie1}), we find:
\begin{equation}
\dot{H}+\frac{3}{2} H^{2}=\frac{\Lambda}{2}\,, \label{frie34}
\end{equation}
where the over-dot denotes derivatives with respect to the cosmic
time. If the vacuum term is negligible, $\Lambda(t) \longrightarrow
0$, then the solution of the above equation reduces to that of the
Einstein de-Sitter model, $H(t)=2/3t$, as it should. Obviously, if
we consider the case of $\Lambda(t)\ne 0$, then it becomes evident
-- see eq.(\ref{frie33}) -- that there is a coupling between the
time-dependent vacuum and matter component. This equation
was first considered by M. Bronstein in a rather old
paper\,\cite{Bronstein33}. Note, that the traditional
$\Lambda=const$ cosmology (or $\CC$CDM model) can be described
directly by the integration of the eq.(\ref{frie34}) [for more
details see section 4.1], but this same equation is also valid for
$\Lambda=\Lambda(t)$, in which case a supplementary equation for the
time evolution of $\Lambda$ is needed in order to unveil the
dynamics of this model. As we have already stated before, the link in
eq.(\ref{frie33}) between $\dot{\rho_m}$ and $\dot{\Lambda}$
is very important because interactions between DM
and DE could provide possible solutions to the cosmological
coincidence problem. This is the reason for which several papers
have been published recently in this area \cite{Zim01} proposing
that the DE 
and DM could be coupled.
Alternatively, one may use a variable $\Lambda$ together with
another entity $X$, that ensures a total self-conserved DE density,
$\rho_{DE}=\rho_{\Lambda}+\rho_X$, such that the ratio
$\rho_{DE}/\rho_m$ remains fairly stable for most of the Universe's
lifetime\,\cite{Grande06}. In both cases a time-dependent DE is
needed.

In the following, we are going to concentrate on models where the
time dependence of $\Lambda$ appears always at the expense of an
interaction with matter. In this context, the corresponding time
evolution equation for the matter density contrast
$D\equiv\delta\rho_m/\rho_m$, in a pressureless fluid, is given
by\,\cite{Arc94}:
\begin{equation}
\label{eq:11}
\ddot{D}+(2H+Q)\dot{D}-\left[\frac{\rho_{m}}{2}-2HQ-\dot{Q}
\right]D=0\,,
\end{equation}
where \be \label{massvac} \rho_{m}=3H^{2}-\Lambda \;\;\;\;\;
Q(t)=-\dot{\Lambda}/\rho_{m} \;\;. \ee It becomes clear, that the
interacting vacuum energy affects the growth factor via the function
$Q(t)$. Of course, in order to solve the above differential
equations we need to define explicitly the functional form of the
$\Lambda(t)$ component. Notice that the approach based on
eq.\,(\ref{eq:11}) effectively implies that the DE
perturbations are negligible (i.e., in this case we set
$\delta\Lambda=0$). This is justified in most cases, specially for
perturbations well inside the sound horizon (of the dark energy
medium) where it behaves very smoothly\,\cite{Grande08,Grande07}.
In fact, one can explicitly derive eq.\,(\ref{eq:11})
starting from the general coupled set of matter and dark energy
perturbations [see equations (17), (25) and (27) of
Ref.\,\cite{Grande08}, which we refrain from repeating here].
Assuming that matter and $\Lambda(t)$ interact as in
eq.(\ref{frie33}), taking the limit where the DE perturbations are
neglected, and assuming also that the produced particles of matter
have negligible velocities with respect to the co-moving observers,
then the aforesaid coupled set of matter and DE perturbation
equations lead to a second order differential equation for $D$ that
boils down to eq.\,(\ref{eq:11}) above. In this way, we can state
that this effective equation follows from the general relativistic
treatment of perturbations, within the aforementioned set of
approximations, whereas in \cite{Arc94} it was originally proven
directly within the Newtonian formalism.

\section{Likelihood Analysis}
In this work, we use a variety of cosmologically relevant observations in order to
constrain the vacuum models explored here (see section 4). These are:
\begin{itemize}
\item {\bf Baryonic Acoustic Oscillations (BAOs):} These are
produced by pressure (acoustic) waves in the photon-baryon plasma in
the early universe, generated by dark matter overdensities. Evidence
of this excess was recently found in the clustering properties of
the luminous SDSS red-galaxies \cite{Eis05} and it can provide a
``standard ruler'' with which we can constraint the dark energy
models. In particular, we use the following estimator\,\cite{Eis05}:
\begin{equation}
A({\bf p})=\frac{\sqrt{\Omega_{m}}}{[z^{2}_{s}E(a_{s})]^{1/3}}
\left[\int_{a_{s}}^{1} \frac{da}{a^{2}E(a)} \right]^{2/3}\,,
\end{equation}
measured from the SDSS data to be $A=0.469\pm 0.017$, with $z_{s}=0.35$
[or $a_{s}=(1+z_{s})^{-1}\simeq 0.75$] and $E(a)\equiv H(a)/H_0$ is the
normalized Hubble flow.
Therefore, the corresponding $\chi^{2}_{\rm BAO}$ function is simply written
\begin{equation}
\chi^{2}_{\rm BAO}({\bf p})=\frac{[A({\bf
p})-0.469]^{2}}{0.017^{2}}\,,
\end{equation}
where ${\bf p}$ is a vector containing the cosmological
parameters that we want to fit.

\item {\bf CMB shift parameter:}
A very accurate and deep geometrical probe of
dark energy is the angular scale of the sound horizon at the last
scattering surface as encoded in the location $l_1^{TT}$ of the
first peak of the Cosmic Microwave Background (CMB) temperature
perturbation spectrum. This probe, called CMB
shift parameter \cite{Bond:1997wr,Nesseris:2006er,Basilakos:2008tk},
is defined as:
\be \label{shift}
R=\sqrt{\Omega_{m}}\int_{a_{ls}}^1 \frac{da}{a^2 E(a)} \;\;. \ee
Note, that the Hubble function $H(a)$ includes also the radiation
component ($\Omega_{r}\simeq 10^{-4}$).
The shift parameter measured from the WMAP 5-years data \cite{komatsu08}
is $R=1.71\pm 0.019$ at $z_{ls}=1090$ [or
$a_{ls}=(1+z_{ls})^{-1}\simeq 9.17\times 10^{-4}$]. In this case,
the $\chi^{2}_{\rm cmb}$ function is given by
\begin{equation}
\chi^{2}_{\rm cmb}({\bf p})=\frac{[R({\bf
p})-1.71]^{2}}{0.019^{2}}\,.
\end{equation}
Note that the measured CMB shift parameter is somewhat model dependent
but mostly to models which are not included in our analysis.
For example, in the case where massive neutrinos are included or when
there is astrongly varying equation of state parameter. The
robustness of the shift parameter was tested and discussed
in \cite{Elgaroy07}.

\item{\bf SNIa distance modulii:}
We additionally utilize the 'Constitution set'
of 397 type Ia supernovae of Hicken et al. \cite{Hic09}. In order
to avoid possible problems related with the local bulk flow,
we use a subsample of the overall sample in which we select
those SNIa with $z>0.023$.
This subsample contains 351 entries.
The corresponding $\chi^{2}_{\rm SNIa}$ function is:
\begin{equation}
\label{chi22}
\chi^{2}_{\rm SNIa}({\bf p})=\sum_{i=1}^{351} \left[ \frac{ {\cal \mu}^{\rm th}
(a_{i},{\bf p})-{\cal \mu}^{\rm obs}(a_{i}) }
{\sigma_{i}} \right]^{2} \;\;.
\end{equation}
where $a_{i}=(1+z_{i})^{-1}$ is the observed scale factor of
the Universe, $z_{i}$ is the observed redshift, ${\cal \mu}$ is the
distance modulus corresponding to flat space:
\be
{\cal \mu}=m-M=5{\rm log}d_{L}+25
\ee
and $d_{L}(a,{\bf p})$ is the luminosity distance
\begin{equation}
d_{L}(a,{\bf p})=\frac{c}{a} \int_{a}^{1} \frac{{\rm
d}y}{y^{2}H(y)}\,,
\end{equation}
Note, that $c$ is the speed of light ($c\equiv 1$ here).
\end{itemize}
We can combine the above cosmologically tests,
using a joint likelihood analysis,
in order to put even more stringent
constraints on the free-parameter space,
according to:
$${\cal L}_{tot}({\bf p})=
{\cal L}_{\rm BAO} \times {\cal L}_{\rm cmb}\times {\cal L}_{\rm
SNIa} $$ or
\be \chi^{2}_{tot}({\bf p})=\chi^{2}_{\rm BAO}+\chi^{2}_{\rm
cmb}+\chi^{2}_{\rm SNIa}\,,
\ee
with the likelihood estimator defined as ${\cal L}_{j}\propto {\rm
exp}[-\chi^{2}_{j}/2]$.
Note, that we sample the $\Omega_{m}$ parameter in steps of 0.01 in
the range [0.1,1]
and that the likelihoods are normalized to their maximum values.
We will report $1\sigma$ uncertainties
of the fitted parameters.
Note that the overall number of
data points used is $N_{tot}=353$ and the degrees of freedom: {\em
  dof}$= N_{tot}-n_{\rm fit}$, with $n_{\rm fit}$ the number of fitted
parameters, which vary for the different models.

\section{Constraints on the time evolving vacuum models in flat space}
As we have already mentioned, the exact nature of a possible time
varying vacuum has yet to be found. A large number of different
phenomenological parameterizations have appeared in the literature treating
the time-dependent $\Lambda(t)$ function. For
example, the authors of\,\cite{Carvalho92} considered
$\Lambda(t)=3\gamma H^{2}$, with the constant $\gamma$ being the
ratio of the vacuum to the sum of vacuum and matter density (see
also \cite{Arc94}), while \cite{Oze87} and \cite{Chen90} proposed a
different ansatz in which $\Lambda(t) \propto a^{-2}$. Also, several
papers, (see for example \cite{Ray07, meng05} and references
therein) have investigated the global dynamical properties of the
universe considering that the vacuum energy density decreases
linearly with the matter energy density, $\Lambda \propto \rho_{m}$.
Carneiro et al. \cite{Carn08, Borg08} used a different pattern in
which the vacuum term is proportional to the Hubble parameter,
$\Lambda(a) \propto H(a)$, while \cite{Bas09a} considered a power
series form in $H$. Attempts to provide a theoretical explanation
for a dynamical $\Lambda(t)$ have also been presented in the
literature using the renormalization group (RG) in quantum field
theory (see \cite{Shap00,SSS04,Grande06,Reuter00} and references
therein). The RG-inspired form for a QFT running vacuum in curved
space-time is

\be \label{RGlaw}\Lambda(H)=n_{0}+n_{2}H^{2}\,, \ee

where $n_{0}$ and $n_{2}$ are constants. This evolution
law can mimic the quintessence or phantom behavior\,\cite{Phantom}
as well as a smoothly transition between the two\,\cite{SS12}.
Notice, that the functional form of eq.(\ref{RGlaw}) has been also used
in \cite{Bas09b}. Such a form is indeed crucially different from
just considering that the vacuum energy is proportional to $H^2$, in
the sense that the former is an ``affine quadratic law'' ($n_0\neq
0$). Remarkably, the structure of eq.(\ref{RGlaw}) can be motivated
from the QFT framework of anomalous induced inflation
(cf.\,\cite{Fossil07} and references therein).
In another vein, the aforementioned possibility that the vacuum energy
could be evolving linearly with $H$ has been motivated theoretically
in the literature through a possible connection of cosmology with
the QCD scale of strong interactions\,\cite{Schu02}. This option,
however, is not what one would expect from re-normalizable QFT in
curved space-time because from general covariance we should rather
expect even powers of $H$, as e.g. in the law (\ref{RGlaw}). There
is, however, the possibility to add non-analytic terms in the
effective action (see \cite{Volov09} for a recent and interesting
attempt in this direction). Such a linear dependence in
$H$ has also been proposed from a possible link of DE with QCD and
the topological structure of the universe\,\cite{Urban09b}. Let us,
however, note that a connection with QCD can also be achieved
through a relaxation mechanism of $\Lambda$ and without using the
hypothesis of linearity in $H$ -- see \cite{FSS09a}.

In this work, we consider a large family of flat vacuum models and
with the aid of the current observational data (see section 2) we
attempt to put stringent constraints on their free parameters. Also, we
investigate thoroughly the time evolution equation of the mass
density contrast in the linear regime as well as the formation of
 galaxy clusters and the evolution of their abundance.
In the following subsections, we briefly
present these cosmological models which trace differently the vacuum
component.

\subsection{The standard $\Lambda$-Cosmology}
Without wanting to appear too pedagogical, we remind the reader of
some basic elements of the concordance $\Lambda$-cosmology in order
to appreciate the differences with the $\Lambda(t)$ models explored
subsequently. The vacuum term in eq.(\ref{frie34}) is constant
and given by $\Lambda=3\Omega_{\Lambda}H^{2}_{0}$. Therefore, it
is a routine to integrate eq.(\ref{frie34}) and obtain the Hubble
function:
\begin{equation}
H(t)=\sqrt{\Omega_{\Lambda}}\;H_{0}
\;{\rm
  coth}\left(\frac{3 H_{0}\sqrt{\Omega_{\Lambda}}}{2}\;t\right)\,,
\label{frie556}
\end{equation}
where $\Omega_{\Lambda}=1-\Omega_{m}$. Note, that $\Omega_{m}$ is the
matter density parameter at the present time.
Using now the definition of the Hubble parameter $H\equiv {\dot a}/a$, the
scale factor of the universe $a(t)$, normalized to unity at the
present epoch, evolves with time as:
\begin{eqnarray}
\label{all}
a(t)=\left(\frac{\Omega_{m}}{\Omega_{\Lambda}}\right)^{1/3}
\sinh^{\frac{2}{3}}\left(\frac{3H_{0}\sqrt{\Omega_{\Lambda}} }{2}\; t\right)
\;\;.
\end{eqnarray}
The cosmic time is related with the scale factor as
\begin{equation}\label{time}
t(a)=\frac{2}{3\sqrt{\Omega_{\Lambda}}H_{0}  } {\rm sinh^{-1}}
\left(\sqrt{ \frac{\Omega_{\Lambda}} {\Omega_{m}}} \;a^{3/2} \right)
\;\;.
\end{equation}

Combining the above equations we can define the normalized
Hubble expansion as a function of the scale factor:
\begin{eqnarray}
\label{hub1} E^{2}(a)=\frac{H^2(a)}{H^2_{0}}=
\Omega_{\Lambda}+\Omega_{m}a^{-3} \;\;\;.
\end{eqnarray}
The inflection point [namely, the point where the
Hubble expansion changes from the decelerating to the
  accelerating regime, $\ddot{a}(t_{I})=0$] takes place at:
\begin{eqnarray}
      \label{infle}
t_{I}=\frac{2}{3\sqrt{\Omega_{\Lambda}}H_{0}} {\rm sinh^{-1}}
\left(\sqrt{ \frac{1} {2}} \right) \;,\;\;
a_{I}=\left[\frac{\Omega_{m}}{2\Omega_{\Lambda}}\right]^{1/3} \;.
\end{eqnarray}
Comparing the concordance model with the observational data we find
that the best fit value, within the $1\sigma$ uncertainty, is
$\Omega_{m}=0.28\pm 0.01$ with $\chi_{tot}^{2}(\Omega_{m})\simeq
431.2$ ({\em dof}=$352$), which is in good agreement with recent studies
\cite{Spergel07,essence, komatsu08, Kowal08, Hic09}. Therefore, the
current age of the universe is $t_{0}\simeq 13.9$Gyr while the
inflection point is located at $t_{I}\simeq 0.52t_{0}$, $a_{I}\simeq
0.58$ (hence at redshift $z_I\simeq 0.72$).
Let us mention that some recent (approximately
model-independent) determinations of this point suggest it to lie at
a more recent time (lower redshift)\,\cite{AVJ09a}, although the
results are still compatible with the $\CC$CDM value within
$2\sigma$.

Finally, solving eq.(\ref{eq:11}) for the $\Lambda$ cosmology
[$Q(t)=-{\dot{\Lambda}}/\rho_{m}=0$], we derive the well known
perturbation growth factor (see \cite{Pee93}):
\be\label{eq24}
D(a)=\frac{5\Omega_{\rm m}
  E(a)}{2}\int^{a}_{0}
\frac{dx}{x^{3}E^{3}(x)} \;\;. \ee

In particular, for $E(a) =\Omega_m^{1/2}\,a^{-3/2}$ it
gives the standard result $D(a)=a$, which corresponds to the
matter dominated epoch, as expected. Notice, that
(\ref{eq24}) is normalized to unity at the present time, $t_0$,
because in our convention $a(t_0)=1$ -- cf. equations
(\ref{all},\ref{time}).

\subsection{The $\Lambda(t)$ model from quantum field theory}

Let us consider the vacuum solution (\ref{RGlaw})
proposed in\,\cite{Shap00,RGTypeIa,SSS04,Fossil07} using the
renormalization group (RG) in quantum field theory (hereafter
$\Lambda_{RG}$ model). We select the coefficients in that equation
as $n_{0}=3\,H_0^2\,(\OL-\gamma)$ and $n_{2}=3\gamma$. Therefore,

\be \label{RGlaw2}\Lambda(H)=\Lambda_0+ 3\gamma\,(H^{2}-H_0^2)\,.
\ee

In this way, the vacuum energy density is normalized to the present
value: $\Lambda_0\equiv\Lambda(H_0)=3\Omega_{\Lambda}H^{2}_{0}$.
Without going into the details of that model, let us recall that
$\gamma$ (called $\nu$ in the above papers) is interpreted in the RG
framework as a ``$\beta$-function'' of QFT in curved-space time,
which determines the running of the cosmological constant. The
predicted value is

\be\label{gammaRG} \gamma=\frac{\sigma}{12\pi}\,\frac{M^2}{M_P^2}\,,
\ee

where $M_P$ is the Planck mass and $M$ is an effective mass
parameter representing the average mass of the heavy particles of
the Grand Unified Theory (GUT) near the Planck scale, after taking
into account their multiplicities.  Since $\sigma=\pm 1$ (depending
on whether bosons or fermions dominate in the loop contributions),
the coefficient $\gamma$ can be positive or negative, but $|\gamma|$
is naturally predicted to be much less than one. For instance, if
GUT fields with masses $M_i$ near $M_P$ do contribute, then
$|\gamma|\lesssim 1/(12\pi)\simeq 2.6\times 10^{-2}$, but we expect
it to be lesser in practice because the usual GUT scales are not
that close to $M_P$. By counting particle multiplicities in a
typical GUT, a natural estimate is the range
$\gamma=10^{-5}-10^{-3}$ (see \,\cite{Fossil07} for details).

We will assume here that $\CC$ evolves in the
form (\ref{RGlaw2}) while it interacts with matter as in
eq.(\ref{frie33}). Alternatively, one may use a variable $\Lambda$
of the form (\ref{RGlaw2}) that interacts with a variable
gravitational constant, $G=G(t)$, such that matter is conserved (see
\,\cite{SSS04,Fossil07}), but we shall not deal with this option
here because we wish to consider only the class of variable $\CC$
models in interaction with matter.

It is important to emphasize that the main motivation for the
evolution law eq.(\ref{RGlaw2}) stems from the general covariance of
the effective action in QFT in curved space-time\,\cite{Shap00,
SSS04, Fossil07} -- for a review, see e.g. \cite{Shapiro:2008sf}.
One expects that deviations from a strictly constant vacuum energy
appear as a result of having a non-trivial external metric that
describes an expanding FLRW background. Since the expansion rate of
this background is $H$, we expect a power series in $H$. However,
for a re-normalizable formulation of the effective action of the
vacuum, only even powers of the expansion rate $H$ can appear, the
leading correction being of ${\cal O}(H^2)$. The next-to-leading
term would be of ${\cal O}(H^4)$, the subsequent one of ${\cal
O}(H^6/M^2)$ etc. At the present time, all of the higher order
corrections are phenomenologically irrelevant compared to the first
curvature correction $\sim {\cal O}(H^2)$ in eq.(\ref{RGlaw2}).
Therefore, it is natural to take just the leading form, as in
eq.(\ref{RGlaw2}).
See, however, \cite{Volov09b} for a possible effect of
$H^4$ terms evaluated at the electroweak crossover scale.

It is interesting to point out that the $\Lambda_{RG}$ model can be
used in different formulations where it helps to alleviate
the cosmic coincidence problem \cite{Grande06,Bas09b}.
It could also have a bearing on the fine tuning problem
\cite{Bas09b}.

From equations (\ref{frie34}) and (\ref{RGlaw2}) we can easily
derive the corresponding Hubble flow as a function of time:
\begin{equation}
\label{frie455} H(t)=H_{0}\,\sqrt{\frac{\OL-\gamma}{1-\gamma}} \;
\coth\left[\frac32\,H_{0}\sqrt{(\OL-\gamma)(1-\gamma)}\;t\right]\,.
\end{equation}
The scale factor of the universe $a(t)$, evolves
with time as
\begin{equation}\label{frie456}
a(t)=a_1\, \sinh^{\frac{2}{3(1-\gamma)}}
\left[\frac32\,H_{0}\sqrt{(\OL-\gamma)(1-\gamma)}\;t\right]\,,
\end{equation}
where
\begin{eqnarray}
\label{all2}
a_{1}=\left(\frac{\Omega_{m}}{\Omega_{\Lambda}-\gamma}\right)^{\frac{1}{3(1-\gamma)}}\,.
\end{eqnarray}
Inverting eq.(\ref{frie456}) we determine the cosmic time as a
function of the scale factor:
\begin{equation}
t(a)=\frac{2}{3\,H_{0}\sqrt{(\OL-\gamma)(1-\gamma)}}\,\sinh^{-1}\left[{\left(\frac{a}{a_1}\right)^{3\,(1-\gamma)/2}}\right]\,.
\label{frie456t}
\end{equation}
The age of the universe is simply given by this expression at $a=1$,
namely at the point $t_0=t(a=1)$. The cosmic time at the inflection
point of the universe evolution [$\ddot{a}(t_{I})=0$] is found to be
\begin{equation}
      \label{inflemod}
t_{I}=\frac{2}{3\,H_{0}\sqrt{(\OL-\gamma)(1-\gamma)}}\,
\sinh^{-1}\left(\sqrt{ \frac{1-3\gamma} {2}} \right)\,,
\end{equation}
and the corresponding value of the scale factor reads
\begin{equation}\label{aIRG}
a_{I}=\left[\frac{(1-3\gamma)\Omega_{m}}{2(\Omega_{\Lambda}-\gamma)}\right]^{\frac{1}{3(1-\gamma)}}\,.
\end{equation}
It becomes clear, that for $\gamma <1/3$ (this parameter is expected
to be very small in QFT) and for the usual values of $\Omega_{\Lambda}$
($\Omega_{\Lambda} > \gamma$) the above inflection point exists.

In practice the condition $\gamma<1/3$ is amply satisfied
in the context of RG-inspired models in which the vacuum energy
evolves as in eq.(\ref{RGlaw2}), the reason being that, in this QFT
context, $\gamma$ appears from eq.\,(\ref{gammaRG}) in which the
highest scale $M$ of the particle masses is expected to lie at (or
below) the Planck scale. Therefore, from QFT we expect
$\gamma\ll 1$ and this is indeed what the comparison with
experimental data confirms (see below). It is thus enlightening to
expand (\ref{aIRG}) in the limit $\gamma\ll 1$:
\begin{eqnarray}\label{aIRG2}
a_{I}=\left(\frac{\Omm}{2\OL}\right)^{1/3}
\left\{1-\gamma\left[1+\frac13\left(\ln\frac{2\OL}{\Omm}-\frac{1}{\OL}\right)\right]\right\}\,.
\end{eqnarray}
Since the coefficient  $\gamma$ in this formula is positive (for the current
values of the cosmological parameters), we see that 
the transition point from deceleration to acceleration occurs earlier in
time than in the concordance $\CC$CDM model (\ref{infle}), whereas
for $\gamma<0$ it occurs at a more recent time.

Let us also consider the behavior of the matter and vacuum
energy densities in this model as a function of the scale factor.
Starting from the conservation law [see eq.\ref{frie33}], and then
trading the time derivatives for derivatives with respect to the
scale factor, we obtain
\begin{equation}\label{Bronstein}
\frac{d\rmm}{da}+\,\frac{3}{a}\,\rmm =-\,\frac{d\CC}{da}\,.
\end{equation}
Using this equation in combination with (\ref{frie33}) and
(\ref{RGlaw2}), we arrive at a simple differential equation for the
matter density,
\begin{equation}\label{rhomRG}
\frac{d\rmm}{da}+\frac{3}{a}\,(1-\gamma)\,{\rmm}=0\,,
\end{equation}
whose trivial integration yields
\begin{equation}\label{mRG}
\rho_m(a) =\rmo\,a^{-3(1-\gamma)}\,.
\end{equation}
Here $\rmo$ is the matter density at the present time ($a=1$). Similarly,
we find
\begin{equation}\label{CRG}
\CC(a)=\CC_0+\frac{\gamma\,\rmo}{1-\gamma}\,\left[a^{-3(1-\gamma)}-1\right]\,.
\end{equation}
Substituting eq.(\ref{frie456}) in the last two equations one may
obtain the explicit time evolution of the matter and vacuum energy
densities, if desired. It is important to emphasize from
eq.\,(\ref{mRG}) that the matter density does no longer evolve as
$\rho_m(a)=\rmo a^{-3}$, as it presents a correction in the
exponent. This is due to the fact that matter is exchanging energy
with the vacuum and this is reflected in the corresponding behavior
of $\Lambda(a)$ in eq.(\ref{CRG}). Substituting eq.(\ref{CRG}) in
eq.(\ref{RGlaw2}), we immediately obtain
\begin{eqnarray}
\label{norm11} E^2(a)&=&
\frac{\OL-\gamma}{1-\gamma}+\frac{\Omm}{1-\gamma}a^{-3\,(1-\gamma)}\nonumber\\
&=& 1+\Omm\, \frac{a^{-3\,(1-\gamma)}-1}{1-\gamma}\,,
\end{eqnarray}
where in the second step we have used the cosmic sum rule for flat
space $\Omm+\OL=1$. Needless to say, eq.\,(\ref{norm11}) can also be
obtained by eliminating the cosmic time from equations
(\ref{frie455}) and (\ref{frie456}), as one can check. It is worth
noting, that the normalized Hubble flow in this model [see
eq.(\ref{norm11})] can be viewed in a similar formulation as
that of the concordance cosmology (for more details see the appendix A) as
follows:
\begin{equation}
\label{anorm11}
E^{2}(a)=\tilde{\Omega}_{\Lambda}+\tilde{\Omega}_{m}a^{-3(1-\gamma)}\,,
\end{equation}
where
\begin{equation}
\label{otran1}
\tilde{\Omega}_{m}=1-\tilde{\Omega}_{\Lambda}=\frac{\Omega_{m}}{1-\gamma} \;\;.
\end{equation}

As expected, for $\gamma\to 0$ ($\tilde{\Omega}_{m} \sim
\Omega_{m}$) all the above equations boil down to the canonical form
within the concordance model -- cf. previous section. Thus, the
traditional $\Lambda$ cosmology is a particular solution of the
$\Lambda_{RG}$ model with $\gamma$ strictly equal to 0.

Comparing the $\Lambda_{RG}$ model with the observational data (we
sample $\gamma \in [-1,0.3]$ in steps of 0.001) we find that the
best fit value is $\Omega_{m}=0.28^{+0.02}_{-0.01}$ (or
$\tilde{\Omega}_{m}\simeq 0.281$) and $\gamma=0.002\pm 0.001$ with
$\chi_{tot}^{2}(\Omega_{m},\gamma)\simeq 431.2$ for {\em dof}$=351$.
We remark that the best fit value that we have obtained for $\gamma$
using the combined set of modern SNIa+BAO+CMB data becomes
significantly smaller than the one obtained in the old analysis of
Ref.\,\cite{RGTypeIa}, where only a limited set of 54 supernovae
data was employed\,\footnote{However, using the current SNIa data
alone our best fit values are in agreement with those found by
\cite{RGTypeIa}.}. In fact, here we have been able to further
restrict the parameter $\gamma$ (called $\nu$ in \cite{RGTypeIa})
and push its value to its natural small range (viz. $\gamma\sim
10^{-3}$ or below) as expected from general QFT
considerations\,\cite{Fossil07} --  see eq.(\ref{gammaRG}).

Being $\gamma$ small and positive, eq.\,(\ref{aIRG2})
predicts that the transition point from deceleration to acceleration
should be slightly earlier in time (hence at larger redshift) as
compared to the standard $\CC$CDM case.

It is interesting to point out that the small $\gamma$ value that we
have obtained from the combined SNIa+BAO+CMB data is nicely
compatible with the result obtained for this parameter from the
analysis of the matter power spectrum of that model, performed
in\,\cite{FSS06} -- \newnewnewtext{see also}
\cite{VelasquezToribio:2009qp}. Here we will analyze also the linear
perturbation regime for the various models. However, rather than
focusing on the power spectrum we will test the implications of this
model on structure formation, through the study of the growth rate
$d\ln D(a)/d\ln a$, the formation of galaxy clusters and the
evolution of their abundances (see sections 5 and 6). In all cases
we need to find the linear matter fluctuation field $D(a)$, which we
shall determine under the assumption of vanishing $\CC$
perturbations. As we have explained in section 2, this is justified
for perturbations well inside the sound horizon and assuming that
the produced matter particles have negligible velocities with
respect to the co-moving observers.

Therefore, we now proceed in an attempt to analytically solve the
differential eq.(\ref{eq:11}) in order to investigate the matter
fluctuation field of the RG model (\ref{RGlaw2}) in the linear
regime. To do so, we change variables from $t$ to a new one
according to the transformation
\be\label{tran11} y={\rm
coth}\left[\frac{3}{2}\,H_{0}\sqrt{(\OL-\gamma)\,(1-\gamma)}\;t\right]
\;\;. \ee
Also from equations (\ref{frie455}, \ref{frie456}) and (\ref{norm11})
we get the following useful relations:
\begin{equation}
\label{pps11}
y=\sqrt{\frac{\beta}{\OL-\gamma}}\,E(a)
\;,\;\;\;\;\;\;
y^{2}-1=\frac{\Omega_{m}}{\OL-\gamma}\;a^{-3\beta}\,,
\end{equation}
where $\beta\equiv 1-\gamma$. Using  (\ref{tran11}) and
(\ref{pps11}) we find, after some algebra,  that equation
(\ref{eq:11}) takes on the form
\be \label{eq:222} 3\beta^2 (y^{2}-1)^{2}D^{''}+f(\beta)
y(y^{2}-1)D^{'}- 2[g(\beta)y^{2}-\psi(\beta)]D=0\,, \ee
where primes denote derivatives with respect to $y$, and
\begin{eqnarray}\label{fgpsi}
&& f(\beta)=2\beta(6\beta-5)\,, \ \
g(\beta)=(2-\beta)(3\beta-2)\,,\nonumber\\
&&\psi(\beta)=\beta(4-3\beta)\,.
\end{eqnarray}
In deriving eq.(\ref{eq:222}) we have substituted the
various terms in (\ref{eq:11}) as a function of the new variable
[see eq.(\ref{pps11})] and the cosmological parameters. For instance, from
equations (\ref{massvac}), (\ref{RGlaw2}), (\ref{mRG}) and (\ref{pps11}), we have
\begin{eqnarray}\label{auxiliar}
\rmm&=& 3\,H_0^2\,(\OL-\gamma)(y^2-1)\nonumber\\
{Q}&=&3\gamma\,H_0\,\sqrt{\frac{\OL-\gamma}{1-\gamma}}\,\,y\nonumber\\
\dot{Q}&=&\frac92\,\gamma\,H_0^2\,(\OL-\gamma)\,(1-y^2)\,.
\end{eqnarray}
Factors of $H_0$ drop at the end of the calculation.
The differential equation (\ref{eq:222}) can be brought into the
standard associated Legendre form by an appropriate transformation.
The growth factor solving this equation reads
\begin{equation}
\label{DRG}
D(y)={\cal C}(y^{2}-1)^{\frac{4-9\beta}{6\beta}} y F\left(\frac{1}{3\beta}+
\frac{1}{2},\frac{3}{2},
\frac{1}{3\beta}+\frac{3}{2},-\frac{1}{y^{2}-1} \right)
\end{equation}
where the quantity $F$ is the hypergeometric function and ${\cal C}$
is a constant (for more details see the appendix B). Inserting
eq.(\ref{pps11}) into eq.(\ref{DRG}) and using (\ref{otran1}), we
finally obtain the growth factor $D(a)$ as a function of the scale
parameter:
\begin{equation}
\label{DRG1}
D(a)=C_{1} a^{\frac{9\beta-4}{2}}E(a)
F\left(\frac{1}{3\beta}+\frac{1}{2},\frac{3}{2},
\frac{1}{3\beta}+\frac{3}{2},-\frac{\tilde{\Omega}_{\Lambda}}{\tilde{\Omega}_{m}}\,\,a^{3\beta}
\right)
\end{equation}
where
\begin{equation}
C_{1}={\cal C} \,\tilde{\Omega}_{\Lambda}^{-1}\left(\frac{\tilde{\Omega}_{m}}
{\tilde{\Omega}_{\Lambda}} \right)^{(4-9\beta)/6\beta} \;\;.
\end{equation}
As a consistency check we note that for $\gamma=0$ ($\beta=1$), the
corresponding normalized to unity growth factor derived from
eq.(\ref{DRG1}) provides the same results as those of the
concordance cosmology (see eq.\ref{eq24}). From this analysis, it
becomes clear that the overall dynamics, predicted by the
$\Lambda_{RG}$ model, extends nicely to that of the usual $\Lambda$
cosmology and connects smoothly to it. For the analysis
of other RG-inspired phenomenological time varying $\CC$ models, see
\,\cite{LixinXu09}.

\subsection{The $\Lambda(t)\propto H^{2}$ model}
We now consider that the vacuum energy density decays as:
$\Lambda=3\gamma H^{2}$ (hereafter $\Lambda_{H_{1}}$ see
\cite{Carvalho92,Arc94}).
This model corresponds in setting
$n_0=0$ in eq.\,(\ref{RGlaw}), and therefore it can be derived as a
particular case of the previous model. In this context, the basic
cosmological equations become
\be\label{Ha} H(t)=\frac{2}{3(1-\gamma)\,t}\,,\;\;\;\;
a(t)=\left[\frac{3(1-\gamma)}{2}H_{0}t\right]^{\frac{2}{3(1-\gamma)}}
\ee
and the normalized Hubble flow is \be E(a)=a^{-3(1-\gamma)/2}  \;\;.
\ee Note, that the constant $\gamma$ lies in the interval $0\le
\gamma < 1$, while the vacuum energy density remains constant
everywhere, which implies that $\Omega_{m}(a)=1-\gamma$.
This is consistent with the cosmic sum rule, since
$\gamma=\OL$ for this model. From eq.(\ref{Ha}) it is obvious
that this model has no inflection point.

If we change the variables from $t$ to $a$ then the time evolution
of the mass density contrast (see eq.\ref{eq:11}) takes the
following form \be
a^{2}D^{''}+\frac{3}{2}a(1+3\gamma)D^{'}-\frac{3}{2}(1+\gamma)(1-3\gamma)D=0
\ee a general solution of which is \be
D(a)=C_{1}a^{1-3\gamma}+C_{2}a^{-3(1+\gamma)/2} \ee where $C_{1}$
and $C_{2}$ are the corresponding constants. Notice, that a growing
mode is present in this scenario if and only if $\gamma <1/3$, which
implies that cosmic structures cannot be formed via gravitational
instability in $\Lambda_{H_{1}}$ models with $\gamma \ge 1/3$. From
a theoretical viewpoint, it is obvious that the $\Lambda_{H_{1}}$
pattern modifies the Einstein de-Sitter model.

We find that the current model is unable to fit the combined
observational data (SNIa+BAO+CMB). Indeed, we find that
$\gamma=0.64^{+0.02}_{-0.01}$ with $\chi^{2}_{SNIa+BAO}\simeq
459.3$. Interestingly, the addition of one more point (the CMB shift
parameter) increases the overall likelihood function by a factor of
$\sim 2.4$ [$\chi_{tot}^{2}(\gamma) \simeq 1104.8$ with
$\gamma=0.95\pm 0.01$]. Using any of the fitted values of
$\gamma$, it then follows from eq.(\ref{Ha}) that for this model the
scale factor evolves as $a\sim t^{\epsilon}$ with $\epsilon>1$.
Under these conditions, this model is free from the horizon
problem. However, unfortunately, it is unable to fit the present
observational data, and moreover combining the latter statistical
result with the $\gamma<1/3$ constrain we conclude that the current
cosmological model is ruled out at high significance level.

\subsection{A power series $\Lambda(t)$ model}
In this case, we parametrize the functional form of $\Lambda(t)$
using a power series expansion in $H$ up to the second order (see
\cite{Bas09a}) and assuming that there is no constant term:
\be\label{PS1}
\Lambda=n_{1}H+n_{2}H^{2} \ee
(hereafter $\Lambda_{PS_{1}}$). Performing the integration of
eq.(\ref{frie34}) we derive the following Hubble function: \be
H(t)=\frac{n_{1}}{\gamma}\frac{{\rm e}^{n_{1}t/2}}{{\rm
e}^{n_{1}t/2}-1} \;\;, \label{frie4} \ee  where \newnewnewtext{we
have defined} $\gamma=3-n_{2}$ \newnewnewtext{and expect} $\gamma$
to remain around $3$ (or $|n_{2}|\ll 1$).
\newnewnewtext{In fact, we do not foresee} that the coefficient $n_2$
could be large (for similar reasons as in section 4.2). Using now
that $H\equiv {\dot a}/a$, the scale factor of the universe $a(t)$,
evolves with time as
\be a(t)=a_{1}\left({\rm e}^{n_{1}t/2}-1 \right)^{2/\gamma} \;\;,
\label{frie44} \ee
where $a_{1}$ is the constant of integration. From these equations,
we can easily write the corresponding Hubble flow as a function of
the scale factor:
\be
H(a)=\frac{n_{1}}{\gamma}\left[1+\left(\frac{a}{a_{1}}\right)^{-\gamma/2}
\right] \;\;. \label{frie5} \ee
Evaluating eq.(\ref{frie5}) at the present time ($a\equiv1$), we
obtain \be n_{1}=\frac{\gamma H_{0}}{1+a_{1}^{\gamma/2}} \;\;.
\label{frie6} \ee Now utilizing equations (\ref{PS1}) and
(\ref{frie6}) and taking into account that the current value of the
vacuum energy density is $\Lambda_{0}=3H_0^2\OL$
($\OL=1-\Omega_{m}$), we arrive at
 \be \label{an1}
a_{1}=\left(\frac{3\Omega_{m}}{\gamma-3\Omega_{m}}\right)^{2/\gamma}\,,
\;\;\;\; n_{1}=H_{0}(\gamma-3\Omega_{m}) \;\;. \ee
Obviously, the above equation implies the following condition:
$\gamma>3\Omega_{m}$ (or $n_{2}<3\OL$).

The normalized to unity, at the present epoch,
scale factor of the universe becomes
\be \label{frie8}
a(t)=\left(\frac{3\Omm}{\gamma-3\Omm}\right)^{2/\gamma}\left[{\rm
e}^{H_0\,(\gamma-3\Omm)\,t/2}-1 \right]^{2/\gamma}
\ee
or
\be
\label{frie9}
t(a)=\frac{2}{H_{0}(\gamma-3\Omega_{m}) }
{\rm ln}\left[\frac{\gamma a^{\gamma/2}E(a)}{3\Omega_{m}} \right]
\ee where \be E(a)=\frac{H(a)}{H_{0}}=1-\frac{3\Omega_{m}}{\gamma}+
\frac{3\Omega_{m}}{\gamma}a^{-\gamma/2} \;\;. \label{frie7} \ee It
is interesting to point here that the current age of the universe
[$a=1$, $E(1)=1$] is \be t_{0}=\frac{2}{H_{0}(\gamma-3\Omega_{m}) }
{\rm ln}\left(\frac{\gamma}{3\Omega_{m}} \right)\,.
\ee

We now study the conditions under which an inflection point exists
in our past, implying an acceleration phase of the scale factor. This
crucial period in the cosmic history corresponds to
$\ddot{a}(t_{I})=0$ and $t_{I}<t_{0}$.
Differentiating twice eq.(\ref{frie8}), we simply have:
\be a_{I}=\left[\frac{3(\gamma-2)\Omega_{m}}{2
(\gamma-3\Omega_{m})}\right]^{2/\gamma}\,, \;\;
t_{I}=\frac{2}{H_{0}(\gamma-3\Omega_{m})} {\rm
ln}(\frac{\gamma}{2})\,, \ee
which implies that the condition for which an inflection point is
present in our past
is $\gamma>2$.

Performing now our statistical analysis we attempt to put constrains
on the free parameters. In particular, we sample $\gamma \in [2,5]$
in steps of 0.01. Thus, the overall likelihood function peaks at
$\Omega_{m}=0.32^{+0.01}_{-0.02}$, $\gamma=3.44 \pm 0.02$ with
$\chi_{tot}^{2}(\Omega_{m},\gamma)=432.7$ ({\em dof}=$351$). Using the latter
cosmological parameters the corresponding current age of the
universe is found to be $t_{0}\sim 14.3$ Gyr while the inflection
point is located at $(a_{I},t_{I})\simeq (0.48,0.42t_{0})$.
This corresponds to a redshift $z_I\simeq 1.08$, which
is substantially higher than in the case of the concordance model.

Following the notations of \cite{Bas09a}, we now derive the growth
factor of fluctuations in the power series model. In particular, we
change variables from $t$ to a new one following the transformation
\be\label{tran1} y={\rm exp}(n_{1}t/2) \;\;{\rm with} \;\; 0<y<1 \;\;. \ee In this
context, using equations (\ref{massvac}), (\ref{PS1}), (\ref{frie5})
and (\ref{tran1}) we obtain
\begin{eqnarray}\label{auxiliar1}
\rmm&=& \frac{n^{2}_{1}y}{\gamma (y-1)^{2}}\nonumber\\
{Q}&=&\frac{n_{1}[(6-\gamma)y-\gamma]}{2\gamma(y-1)}\nonumber\\
\dot{Q}&=&\frac{n^{2}_{1}(\gamma-3)y}{2\gamma(y-1)^{2}}\;\;,
\end{eqnarray}
(for $n_{1}$ see eq.\ref{an1}).

We can now re-write eq.(\ref{eq:11}) as:
\be\label{eq:22} \gamma^{2} y(y-1)^{2}D^{''}+2\gamma
(y-1)(5y-\gamma)D^{'}-2(6-\gamma)(\gamma-2y)D=0 \ee where prime
denotes derivatives with respect to $y$. Notice, that this
variable is related with the scale factor as \be
y=1+\frac{\gamma-3\Omm}{3\Omm}\,a^{\gamma/2} \,.
\ee
We find that eq.(\ref{eq:22}) has a decaying solution of the
form $D_{1}(y)=(y-1)^{(\gamma-6)/\gamma}\sim a^{(\gamma-6)/2}$ for $\gamma<8$.
Thus, the corresponding growing mode of eq.(\ref{eq:22}) is \be
\label{eqff2} D(a)=C a^{(\gamma-6)/2}\int_{0}^{a} \frac{{\rm
d}x}{x^{\gamma/2}E^{2}(x)} \ee where \be
C=\frac{\gamma}{2}\tilde{\Omega}_{m}^{2}\left(\frac{\tilde{\Omega}_{\Lambda}}
{\tilde{\Omega}_{m}} \right)^{(2\gamma-4)/\gamma} \ee with \be
\label{otran2}
\tilde{\Omega}_{m}=1-\tilde{\Omega}_{\Lambda}=\frac{3\Omega_{m}}{\gamma}
\;\;. \ee It is interesting to mention that, the normalized Hubble
flow [see eq.(\ref{frie7}) and appendix A] can be cast as:
\be \label{frie77}
E(a)=\tilde{\Omega}_{\Lambda}+\tilde{\Omega}_{m}a^{-\gamma/2}\;\;.
\ee It becomes clear that for $\gamma \to 3$ we have
$\Omega_{m} \sim \tilde{\Omega}_{m}$. Using our best fit values we
find $\tilde{\Omega}_{m}\simeq 0.28$. Finally, in the limit of
$n_1\to 0$ (or $n_2\to 3\tilde{\gamma}$; $\tilde{\gamma}$ is a
constant) the current cosmological pattern tends to the
$\Lambda_{H_{1}}$ model (see section 4.3) as it should
[$\Lambda(a)\sim 3{\tilde{\gamma}} H^{2}$, $E(a)\sim
a^{-3(1-\tilde{\gamma})/2}$ and $D(a)\sim a^{1-3\tilde{\gamma}}$].

\subsection{The $\Lambda\propto H $ model}

As we have previously mentioned, different authors
\cite{Schu02,Volov09,Urban09b,Volov09b} have tried to provide some
fundamental reason for the possibility that the vacuum term could be
proportional to the Hubble parameter, $\Lambda(a) \propto H$.
Note, that this kind of cosmological model (hereafter
$\Lambda_{PS_{2}}$) is a particular case of the power series model
studied in the previous section by setting $n_{2}=0$ in
eq.(\ref{PS1}):
\be\label{PS2} \Lambda=n_{1}H\,. \ee
In this case, the normalized Hubble flow simply reads as in
(\ref{frie77}), but using the original (untilded) cosmological
parameters and choosing $\gamma=3$:
\be \label{frie77b} E(a)=\OL+\Omm\,a^{-3/2}\;\;. \ee
One of the merits of the $\Lambda_{PS_{2}}$ vacuum model is that it
contains only one free parameter, $n_1$. Obviously, this parameter
is determined to be
\begin{equation}\label{n1PS2}
n_1=3\,H_0\,\OL\,.
\end{equation}
Therefore, since we use the prior $H_{0}=70.5$ km/sec/Mpc
\cite{komatsu08}, the free parameter is actually $\OL$ (or,
equivalently, $\Omm=1-\OL$). This single parameter is the same as in
the standard flat $\CC$CDM model, except that the Hubble rate in
(\ref{frie77b}), after squaring it, compares very differently with
eq. (\ref{hub1}). Needless to say, this could make a dramatic
difference when we try to fit the the combined SNIa+BAO+CMB data
with the $\Lambda_{PS_{2}}$ model.

Indeed, if we now marginalize the results of the previous section
over $\gamma=3$, the joint likelihood analysis provides a
best fit value of $\Omega_{m}=0.34 \pm 0.01$, which is significantly
larger than in the $\CC$CDM case, but with a poor quality fit:
$\chi_{tot}^{2}(\Omega_{m})\simeq 513.6$ for {\em dof}$=352$
\cite{Bas09a}.

This simply means that the functional form
(\ref{frie77b}) is unable to fit the observational data
simultaneously at low and high redshifts. We confirm this point by
using the CMB shift parameter only, finding that the corresponding
likelihood function peaks at $\Omega_{m}\simeq 0.80$.
This value is $\sim 2.5$ times larger than that provided by the
SNIa+BAO solution $\Omega_{m} \simeq 0.32$.
This fact alone suggests that
in spite of the various adduced motivations in the literature for
the class of models $\CC\propto H$, they are unfortunately unable to
provide a quality fit of the basic cosmological data in all the
relevant redshift ranges.

We note that although our combined SNIa+BAO+CMB likelihood analysis
provides a similar $\Omega_{m}$ value to that found in 
\cite{Carn08}, it has a quite large reduced $\chi^{2}$ ($\simeq 1.46$),
which is in contrast with Carneiro et al. who concluded that the 
$\CC\propto H$ model was compatible with the data they
used. However, this conclusion immediately paled after performing
the study of the matter power spectrum of such model, which turned
out to be highly unfavorable owing to a significant late-time
depletion of power as compared to the standard $\CC$CDM model
\,\cite{Borg08}. We find that this is borne out by the present
study; in fact, the growth rate of galaxy clustering becomes frozen
at late times for that model (as we will see in section
\ref{linearfield}). The analysis of the fluctuation field $D(a)$ is
performed exactly as in section 4.4 in the limit $\gamma\to 3$.

\subsection{A power law $\Lambda(t)$ model}
In this phenomenological scenario, we generalize the ideology of
\cite{Oze87,Chen90} in which the vacuum energy density evolves as
$\Lambda(a)\propto a^{-2}$. In the present work, the vacuum energy
is taken to evolve with an arbitrary power of the scale factor:
\be \label{powerlaw} \Lambda(a)=3\gamma(3-n)a^{-n} \ee
(hereafter $\Lambda_{n}$ model; see also \cite{Silv94}). The
corresponding Hubble flow as a function of the scale factor follows
from solving eq.\,(\ref{frie34}) with $\CC$ given as in
eq.(\ref{powerlaw}). Trading the cosmic time variable for the
scale factor, the differential equation (\ref{frie34}) can be seen as
\begin{equation}\label{4ta}
\frac{dH^2}{da}+\frac{3}{a}\,H^2=3\gamma\,(3-n)\,a^{-n-1}\,.
\end{equation}
The corresponding solution satisfying the boundary condition
$H(a=1)=H_0$ reads as follows:
\be \label{HUBLAW} H^{2}(a)=(H_0^2-3\gamma)a^{-3}+3\gamma a^{-n}\,.
\ee
Using now the following parametrization,
\begin{equation}
\label{otran3}
\tilde{\Omega}_{\Lambda}=1-\tilde{\Omega}_{m}=\frac{3\gamma}{H^{2}_{0}} \;\;,
\end{equation}
we simply derive
\be\label{Eapow}
E^{2}(a)=\tilde{\Omega}_{m}a^{-3}+\tilde{\Omega}_{\Lambda}a^{-n}
\;\;. \ee
Evidently, if we parameterize the constant $n$ according to
$n=3(1+w)$ then the current $\Lambda_{n}$ cosmological model can be
viewed as a classical {\em quintessence} model ($P_{Q}=w\rho_{Q}$)
with $w<-1/3$ (or $n<2$), as far as the global dynamics is concerned
\cite{Silv94}), despite the fact that the two models have a
different equation of state parameter. On the other hand, utilizing
eq.(\ref{powerlaw}) at the present epoch
[$\Lambda_{0}=3\gamma(3-n)$] and taking into account that the
current value of the vacuum energy density is
$\Lambda_{0}=3H_0^2\OL$ we obtain
\begin{equation}\label{gammapow}
\gamma=\frac{\OL\,H^{2}_{0}}{3-n}\,,
\end{equation}
and
\begin{equation}
\label{otran4}
\tilde{\Omega}_{\Lambda}=1-\tilde{\Omega}_{m}=\frac{3\Omega_{\Lambda}}{3-n} \;\;.
\end{equation}
Obviously, for $n \to 0$ we get
$\tilde{\Omega}_{\Lambda} \sim \Omega_{\Lambda}$
(or $\tilde{\Omega}_{m} \sim \Omega_{m}$) as we should.

It is worthwhile to compute the
evolution of the matter density as a function of the scale factor.
\newnewtext{From} (\ref{frie1}), (\ref{powerlaw}) and (\ref{HUBLAW}) we find
\begin{equation}\label{rhom}
\rho_m(a)=3\left(H_0^2-3\gamma\right)\,a^{-3}+3n\,\gamma a^{-n}\,.
\end{equation}
We remark that, in spite of the simple form of the Hubble rate (\ref{Eapow})
in terms of the formal parameters
$(\tilde{\Omega}_{m},\tilde{\Omega}_{\Lambda})$ (see Appendix A),
the matter density is a mixture of the canonical $a^{-3}$ component
and the new $a^{-n}$ component associated to the vacuum energy. This
was expected from the fact that matter and vacuum energy are in
interaction in this model. The physical parameters $(\Omm,\OL)$ are
related in the standard manner. Indeed, since
$\rho_{m0}=\rho_m(a=1)$, we have from (\ref{rhom})
\begin{equation}\label{Ompow}
\Omm=\frac{\rho_{m0}}{3H_0^2}=1-\frac{\gamma (3-n)}{H_0^2}=1-\OL\,,
\end{equation}
which is consistent with the result found above for $\OL$.

Comparing the $\Lambda_{n}$ model with the observational data (we
sample $n \in [-0.2,2]$ in steps of 0.01) we find that the best fit
values are $\Omega_{m}=0.29^{+0.01}_{-0.02}$ (or
$\tilde{\Omega}_{m}\simeq 0.28$) and $n=-0.06 \pm 0.04$ with
$\chi_{tot}^{2}(\Omega_{m},n)\simeq 431.2$ ({\em dof}=$351$).
The standard $\CC$CDM case is obtained from this model for $n=0$, and
the best fit value is indeed close to it, but the fact that $n$
approaches $0$ from below (implying $w\simeq -1.02$) means that the
preferred situation is slightly tilted into the phantom domain.
The inflection point for this model can be easily
computed without resorting to the time dependence as follows.
Requiring that the deceleration parameter
\begin{equation}\label{qparam}
q=-\frac{\ddot{a}a}{\dot{a}^2}=-1-\frac{a}{2\,H^2(a)}\,\frac{dH^2}{da}
\end{equation}
vanishes, we find that $a\,dH^2/da+2H^2=0$, and hence $a_I$ is the
root of this equation. Using (\ref{HUBLAW}) and (\ref{gammapow}), we
are immediately led to
\begin{equation}
a_{I}=\left[\frac{3\Omega_{m}-n}{3(2-n)\Omega_{\Lambda}}\right]^{1/(3-n)}\,,
\end{equation}
which implies that $n<2$ for $\Omega_{m} \in (0,1)$. Using our best
fit values, we find $a_{I}\simeq 0.60$ ($z_I\simeq 0.66$), while the
age of the universe is $t_{0}\sim 13.6$Gyr. We remark
that, contrary to the previous models, the power law model predicts
a transition point from deceleration to acceleration located at a
time more recent than the concordance model.

Finally, if we change the variables from $t$ to $a$ then
the evolution of the mass density contrast
(see eq.\ref{eq:11}) becomes:
\be
\label{Dzzl}
a^{2}H^{2}D^{''}+a[3H^{2}+aHH^{'}+HQ]D^{'}-G(a)D=0
\ee
where
\be
G(a)=\frac{\rho_{m}}{2}-2HQ-aHQ^{'} \;\; .
\ee
Due to the fact that the best fit value $n=-0.06$ is relatively
small, we can neglect from eq.(\ref{Dzzl}) the corresponding
high order terms ($n^{2}$, $n^{3}$, $n\Omega_{m}\Omega_{\Lambda}$, etc). Thus,
we obtain the following accurate formula to the growth factor
(see also \cite{Silv94})
\be
D(a)\simeq a\;F\left(-\frac{1}{3w},\frac{w-1}{2w},1-\frac{5}{6w},
-\frac{\tilde{\Omega}_{\Lambda}a^{-3w}}{\tilde{\Omega}_{m}}
 \right)
\ee where $w=-1+n/3$ and $F$ is the hypergeometric function.
Notice that $D(a)\sim a$ for $\OL\to 0$, as expected.

\section{Linear growth rate for the time varying vacuum models }
\label{linearfield}

In the upper panel of figure 1 we present the growth factor
evolution, for the different time varying vacuum models,
presented in this work, as a function
of redshift, $D(z)$ [$z=a^{-1}-1$].
Notice, that the growth factors are normalized to unity at the
present time ($z=0$). We find that for $z\le 0.5$ the
$\Lambda_{PS_{1}}$ and $\Lambda_{PS_{2}}$ growth factors reach a
plateau, implying that the matter fluctuations are effectively
frozen, as we already advanced in section 4.5. Also, it is obvious
that the growth factors for the latter vacuum models are much
greater with respect to the other 3 models, $\Lambda$,
$\Lambda_{RG}$ and $\Lambda_{n}$. Therefore, it is expected that
this difference among the above cosmological models will affect also
the predictions related with the formation of the cosmic structures
(see next section), due to the fact that they trace differently the
evolution of the matter fluctuation field.

\begin{figure}[t]
\mbox{\epsfxsize=10cm \epsffile{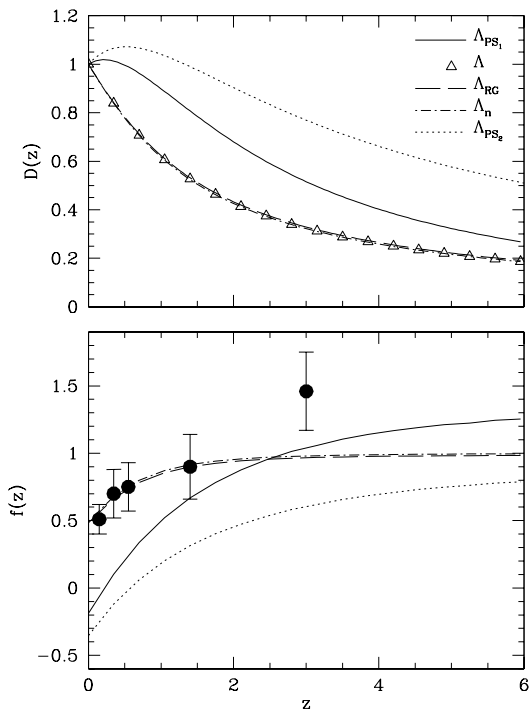}} \caption{{\it Upper Panel:}
The evolution of the growth factor. The lines correspond to
$\Lambda_{PS_{1}}$ (solid), $\Lambda_{RG}$ (dashed), $\Lambda_{n}$
(dot-dashed) and $\Lambda_{PS_{2}}$ (dot). Note, that open triangles
correspond to the traditional $\Lambda$ cosmology. {\it Bottom
Panel:} Comparison of the observed (solid circles\,\cite{Ness08},
(see Table I) and theoretical evolution of the growth
rate of clustering
$f(z)$. We do not plot the growth rate of the $\Lambda$ cosmology in
order to avoid confusion.}
\end{figure}
\begin{table}[ht]
\caption[]{Data of the growth rate of clustering \cite{Ness08}. The
correspondence of the columns is as follows: redshift, observed
growth rate and references.} \tabcolsep 4.5pt
\begin{tabular}{ccc} \hline \hline
z& $f_{obs}$ & Refs. \\ \hline
0.15 & $0.51\pm 0.11$& \cite{Verde02,Hawk03}\\
0.35 & $0.70\pm 0.18$& \cite{Teg06} \\
0.55 & $0.75\pm 0.18$& \cite{Ross07}\\
1.40 & $0.90\pm 0.24$& \cite{daAng08}\\
3.00 & $1.46\pm 0.29$& \cite{McDon05}\\
\end{tabular}
\end{table}
\begin{table}[ht]
\caption[]{The reduced $\chi^2$ values ($\chi^{2}_{min}/{\rm df}$) and
KS probabilities comparing the growth rate of clustering between
data and vacuum model expectations.}
\tabcolsep 2.5pt
\begin{tabular}{ccccc} \hline \hline
Model& $\chi^{2}_{min}/5$ & ${\cal P_{KS}}$ & $\chi^{2}_{min}/4(z<3)$ & ${\cal P_{KS}}(z<3)$ \\ \hline
$\Lambda$ & 0.63 & 1.0 & 0.10&0.997\\
$\Lambda_{RG}$ & 0.64 & 1.0 & 0.10& 0.997\\
$\Lambda_{PS_{1}}$ & 9.7& 0.210 & 11.6& 0.107 \\
$\Lambda_{PS_{2}}$ & 20& 0.036 & 22.9 & 0.011\\
$\Lambda_{n}$ & 0.63& 1.0 & 0.10 & 0.997\\
\end{tabular}
\end{table}
We would like to end this section with a discussion on the evolution
of the well known indicator of clustering, namely the growth rate
\cite{Pee93}:
\be
f(a)\equiv \frac{d{\rm ln}D}{d{\rm ln}a} \;.
\ee
From
the known analytical form of the growth factor $D(a)$ for the
current vacuum models\,\footnote{The formula $\frac{d}{dx}F\left(a,b,c,x\right)=\frac{ab}{c}F\left(a+1,b+1,c+1,x\right)$
for differentiating the hypergeometric function is used to compute
$f(z)$ from $D(z)$.}, in the bottom panel of Fig.\,1 we display the
predicted growth rate:
\be
f(z)=-(1+z)\frac{d{\rm ln}D}{dz}\;,
\ee
together with the
observed $f_{obs}(z)$ [filled symbols] using the recent results of
the 2dF and SDSS galaxies \cite{Ness08}. In Table I, we
quote the precise numerical values of the data points, with the
corresponding error bars. We compare the growth rate of clustering
between data and models via a $\chi^{2}$ minimization and a
Kolmogorov-Smirnov (KS) statistical test respectively. For the model
predictions we use the best fitted valued for the parameters
($\gamma$ or $n$) obtained in section 4 for each model. We then
compute the corresponding consistency between models and data
($\chi^{2}_{min}/5$ and ${\cal P}_{KS}$) and place these results in
Table II. Obviously, in the case of $\Lambda$, $\Lambda_{RG}$ and
$\Lambda_{n}$ models we find good consistency ($\chi^{2}_{min}/5
\simeq 0.64$ and ${\cal P}_{KS}\simeq 1$). Therefore, it is apparent
that the power series  $\Lambda_{PS_{1}}$ and $\Lambda_{PS_{2}}$
vacuum models (and specially the latter, which is the
model where $\CC$ evolves linearly with $H$) fail to fit the data
($\chi^{2}_{min}/5 \simeq 20$ and ${\cal P}_{KS}\simeq 0.036$ for
$\Lambda_{PS_{2}}$). In this framework, close to the present epoch
$z\le 0.5$,  the $\Lambda_{PS_{1}}$ vacuum model is unable to fit
the data ($\chi^{2}_{min}/5 \simeq 9.7$ and ${\cal P}_{KS}\simeq
0.21$) because the matter fluctuation field is frozen and thus the
growth rate of clustering effectively overs. However, increasing the
free parameter $\gamma$ by a factor of $\sim 1.2$ ($\gamma \simeq
4.1$) the $\Lambda_{PS_{1}}$ model appears to fit the $f_{obs}(z)$
data ($\chi^{2}_{min}/5\simeq 1.2$ and ${\cal P}_{KS}\simeq 0.99$).
Note, that in the last two rows of Table II, we list the
corresponding
results by excluding from the statistical analysis the observed
growth rate of clustering at $z=3$ (see Table I \cite{McDon05}).

\section{The formation and evolution of collapsed structures}
In this section we study the cluster formation
processes by using the usual Press-Schecther formalism \cite{press}, which
studies the the behavior of the matter perturbations assuming
a Gaussian random field background, but applied also
within the framework of the vacuum models studied in this work.

\begin{figure}[ht]
\mbox{\epsfxsize=10cm \epsffile{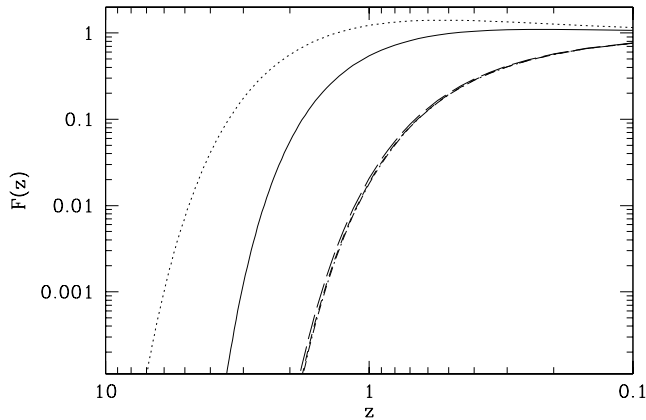}} \caption{The predicted
fractional rate of cluster formation as a function of redshift for
the current cosmological models (using $\sigma_{8}=0.80$).
The meaning of the various lines is as in Fig.\,1.}
\end{figure}

We wish here to estimate, within the different vacuum models,
 the fractional rate of cluster formation (see \cite{Peeb84,
   Rich92}). In particular, these studies introduce a methodology which computes the
rate at which mass joins virialized halos (such as galaxy clusters), which grow from small
initial perturbations in the universe,
with matter fluctuations, $\delta$, greater than a critical value $\delta_{c}$.

Assuming that the density fluctuation field, smoothed
at the scale $R$ (corresponding to a mass scale of $M=4\pi\bar{\rho} R^3/3$, with $\bar{\rho}$
the mean background mass density of the Universe), is normally distributed
with zero mean, then the probability that the field will have a value
$\delta$ at any given point in space is:
\be\label{eq:88}
{\cal P}(\delta, z) =\frac{1}{\sqrt{2\pi}\sigma(R,z)}
{\rm exp}\left[-\frac{\delta^{2}}{2\sigma^{2}(R,z)} \right]\;,
\ee
where the variance of the Gaussian field, $\sigma^2(R,z)$, is given by:
\be
\sigma^2(R,z) = \frac{1}{2 \pi^2} \int_0^{\infty} k^2 P(k,z) W^2(kR) dk
\ee
with $P(k,z)$ the power spectrum of density fluctuations
which
evolves according to $P(k,z)=P(k,0) D^2(z)$, with $D(z)$ the growing
mode of the density fluctuations evolution, normalized such that
$D(0)=1$.
Finally, $W(kR)$ is the top-hat
smoothing kernel, given in Fourier-space by:
\be
W(kR)=\frac{3}{(kR)^3} [\sin(kR)-kR \cos(kR)] \;.
\ee
We can now estimate what fraction of the Universe, at some reference
redshift $z$
and above some mass threshold $M$, has collapsed to form bound
structures.
To this end we need to integrate the probability function
given by eq.(\ref{eq:88}) over all regions that at
some prior redshift had overdensities which by the reference redshift
have increased to above the critical value, $\delta_c(z)$, which in an
Einstein-deSitter universe is $\simeq 1.686$ and varies slightly for
different values of $\Omega_m$ \cite{LaceyCole}.
Therefore, this fraction
is given by \cite{Rich92}:
\be
{\cal F}(M,z)=\int_{\delta_{c(z)}}^{\infty} {\cal P}(\delta, z) d\delta \;,
\ee
and performing the above integration,
parametrizing the rms mass fluctuation amplitude at
$R=8 \; h^{-1}$ Mpc, which can be expressed as a function of redshift as
$\sigma(M,z)=\sigma_{8}(z)=D(z)\sigma_{8}$, we obtain:
\be
\label{eq:89}
{\cal F}(z)=\frac{1}{2} \left[1-{\rm erf}
\left( \frac{\delta_{c}}{\sqrt{2} \sigma_{8}(z)} \right) \right] \;\;.
\ee
Obviously the above generic form of eq.(\ref{eq:89}) depends on the
choice of the background cosmology and the power-spectrum
normalization, which we take to be the
WMAP5 result of $\sigma_{8} \simeq 0.80$ \cite{komatsu08}.

The final step is to normalize the above probability
to give the fraction of mass in bound structures
which have
already collapsed by the epoch $z$, divided by the corresponding fraction
in structures which have collapsed at the present epoch ($z=0$),
\be \tilde{F}(z)={\cal F}(z)/{\cal F}(0) \;.\ee
In Fig.\,2 we present in a logarithmic scale the
behavior of normalized structure formation rate as a function of
redshift for the present vacuum models.

In the context of the power series vacuum energy models
(see $\Lambda_{PS_{1}}$ solid line and $\Lambda_{PS_{2}}$ dot line
in Fig.\,2), we find that prior to $z\sim 0.5$ the cluster formation
has effectively terminated due to the fact that the matter fluctuation field,
$D(z)$, effectively freezes. Also, the large amplitude of the
$\Lambda_{PS_{1}}$ and $\Lambda_{PS_{2}}$ fluctuation field
(Fig.\,1) implies that in these models galaxy clusters
appear to form earlier ($z\ge 4$) with respect to the $\Lambda$,
$\Lambda_{RG}$ and $\Lambda_{n}$ (dashed lines) vacuum models.
Indeed, for the latter cosmological models we find that galaxy
clusters formed typically at $z\sim 2$. Finally, it is worth
noticing that for a higher value of $\sigma_{8} (>0.80)$, the corresponding
cluster formation rate moves to higher redshifts and obviously, the opposite
situation is true for $\sigma_{8}<0.80$.

\subsection{The halo abundance and its evolution}
From the previously presented Press-Schecther formalism (hereafter PSc),
using the fraction of the universe, $F(M,z)$, that has
collapsed by some redshift in halos above some mass $M$, we can estimate
the number density of halos, $n(M,z)$, with masses with a range $(M, M+\delta M)$, by the following:
\be
n(M,z) dM= \frac{\partial {\cal F}(M,z)}{\partial M} \frac{{\bar \rho}}{M} dM\;.
\ee
Performing the differentiation and after some algebra we derive the
following:
\begin{eqnarray}
n(M,z) dM
&=& -\frac{\bar{\rho}}{M} \left(\frac{1}{\sigma}
\frac{d \sigma}{d M}\right)
f_{\rm PSc}(\sigma) dM \nonumber\\
&=& \frac{\bar{\rho}}{M} \frac{d{\rm \ln}\sigma^{-1}}{dM}
f_{\rm PSc}(\sigma) dM
\end{eqnarray}
where $f_{\rm PSc}(\sigma)=\sqrt(2/\pi) (\delta_c/\sigma)
\exp(-\delta_c^2/2\sigma^2)$. Note that within this approach all the
mass is locked inside halos, according to the normalization
constraint:
\be
\int_{-\infty}^{+\infty} f_{\rm PSc}(\sigma) d{\rm \ln}\sigma^{-1} = 1\;.
\ee
Although the above (Press-Schecther) formulation was shown to give a good rough
approximation to the expectations provided by numerical simulations,
it was later found to overpredict/underpredict the number of low/high mass halos at the
present epoch (eg. \cite{Jenk01} and references therein). There is a
large number of works providing better fitting
functions of $f(\sigma)$, which are mostly based on a phenomenological approach
(see \cite{Reed} and references therein).

From the halo mass function we can now derive an observable quantity
which is the redshift distribution of clusters, ${\cal N}(z)$, within some
determined mass range, say $M_1\le M/M_{\odot}\le M_2$.
This can be estimated by integrating, in mass, the expected
differential halo mass function, $n(M,z)$, according to:
\be
{\cal N}(z)=\frac{dV}{dz}\;\int_{M_{1}}^{M_{2}} n(M,z)dM
\ee
where $dV/dz$ is the comoving volume element, which in a flat
universe takes the form:
\be
\frac{dV}{dz}
=4\pi r^{2}(z)\frac{dr}{dz}(z)
\ee
with $r(z)$ the comoving radial distance out to redshift $z$:
\be
r(z)=\frac{c}{H_{0}} \int_{0}^{z} \frac{dx}{E(x)} \;.
\ee We use the halo mass function of Reed et al. \cite{Reed},
integrating for cluster-size halo masses, ie.,  between: $10^{13.4}
< M/M_{\odot} <10^{16}$. In Fig.\,3 we show the integral halo mass
function, $n(>M)$, for the models discussed previously and at two
different redshifts. We use the WMAP5 normalization of the
power-spectrum, ie., $\sigma_8\simeq 0.8$. Note that the classical
$\Lambda$, the $\Lambda_{RG}$ and $\Lambda_n$ models provide
indistinguishable halo mass functions, with those corresponding
to the $\Lambda_{PS2}$ and $\Lambda_{PS1}$
models being way off (we plot results only of the former model).
\begin{figure}[ht]
\mbox{\epsfxsize=8.5cm \epsffile{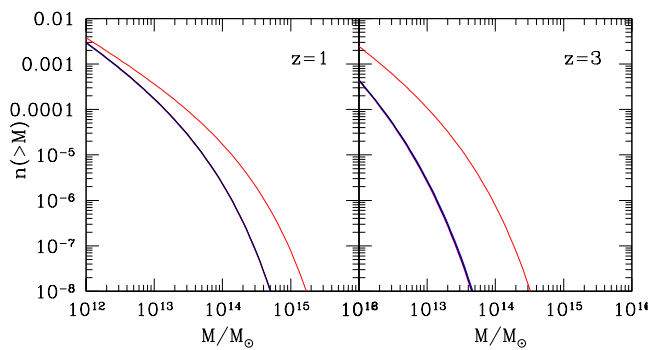}}
\caption{The halo mass function at two different redshifts.
The different models are represented
  by the same line-types as in the previous figures (which however
  fall on each other and thus are indistinguishable), but only the
  $\Lambda_{PS2}$ model (upper curve) provides a significantly
different $n(M,z)$ that the other models.}
\end{figure}

In Fig.\,4 we show theoretically expected cluster redshift
distribution for the three closely resembling models, ie., the
standard $\Lambda$ model, the $\Lambda_{RG}$ and $\Lambda_n$ models
(left panel) and the fractional difference between the first
(constant $\Lambda$) and each of the other two models (right panel).
The expected differences are small at low redshifts, but become
gradually larger for $z\gtrsim 2$, reaching variations of up to
$\sim 100\%$ at $z\sim 5$.
\begin{figure}[ht]
\mbox{\epsfxsize=9cm \epsffile{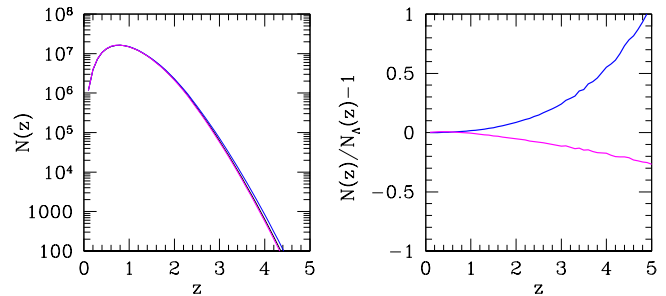}} \caption{The expected cluster
redshift distribution (left panel) and
  the corresponding \newnewnewtext{fractional difference (right panel) of}
the $\Lambda_{RG}$ (upper blue curve) and $\Lambda_n$ (lower magenta
curve) models with respect to the standard $\Lambda$ model.}
\end{figure}
To investigate
how realistic it would be to detect such differences,
we provide below the expectations for two realistic cluster surveys covering
the hole sky (although a more realistic case would be to consider a
solid angle of the order of $\sim 4 \times 10^3$ square degrees):

\noindent
(a) an X-ray survey down to a flux of: $f_{\rm lim}=3\times 10^{-14}$ ergs s$^{-1}$
cm$^{-2}$, as that expected from the future {\tt eROSITA} X-ray satellite,
and

\noindent
(b) a Sunayev-Zeldovich (SZ) survey with a limiting flux
density at $\nu_0=150$ GHz of $f_{\nu_0, {\rm lim}}=5$ mJy
(as expected from the survey of the Southern Polar Telescope).

To realize the predictions of the first survey
we use the relation between halo mass and bolometric X-ray
luminosity, as a function of redshift, provided in \cite{Fedeli}, ie:
\be
L(M,z)=3.087 \times 10^{44} \left[\frac{M_{h} h E(z)}{10^{15}
    M_{\odot}} \right]^{1.554} h^{-2} \; {\rm erg s^{-1}} \;.
\ee
The limiting halo mass that can be observed at redshift $z$ is then
found by inserting in the above equation the limiting luminosity, given
by: $L=4 \pi d_L^2 f_{\rm lim}$, with $d_L$ the luminosity distance
corresponding to the redshift $z$.

The predictions of the second survey can be realized using again the
relation between limiting flux and halo mass from \cite{Fedeli}:
\be
f_{\nu_0, {\rm lim}}= \frac{2.592 \times 10^{8} {\rm mJy}}{d_{A}^{2}(z)}
\left(\frac{M}{10^{15} M_{\odot}}\right)^{1.876} E^{2/3}(z) \;
\ee
where $d_A(z) \equiv d_L/(1+z)^2$ is the angular diameter distance out
to redshift $z$.

In Fig.\,5 we present the expected redshift distribution and the
fractional difference between the different models, similarly to
Fig.\,4, but now for the realistic case of the X-ray survey.
Similarly, in Fig.\,6 we present the corresponding results for the
case of the SZ survey.
\begin{figure}[ht]
\mbox{\epsfxsize=9cm \epsffile{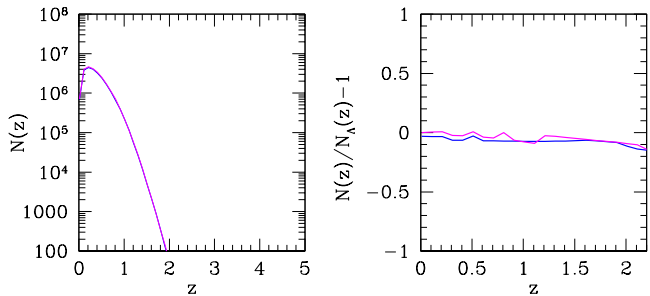}}
\caption{The expected cluster redshift distribution (left panel) and
  the corresponding \newnewnewtext{fractional difference} (right
  panel) of the $\Lambda_{RG}$ and $\Lambda_n$ models with respect
  to the standard $\Lambda$ model for the case of a realistic (future) X-ray
  survey with a flux limit of $10^{-14}$ erg s$^{-1}$ cm$^{-2}$.}
\end{figure}
\begin{figure}[ht]
\mbox{\epsfxsize=9cm \epsffile{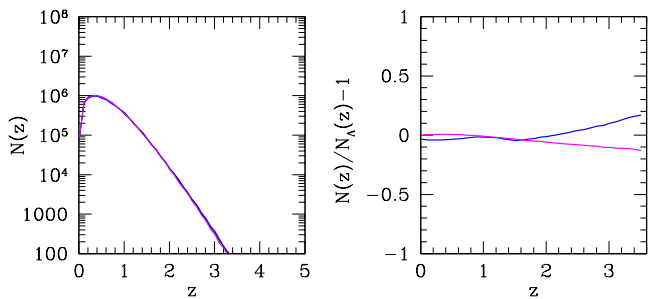}}
\caption{The expected cluster redshift distribution (left panel) and
  the corresponding \newnewnewtext{fractional difference} (right
  panel) of the $\Lambda_{RG}$ (upper blue curve)
and $\Lambda_n$ (lower magenta curve) models
  with respect to the standard $\Lambda$ model for the case of a realistic
  (future) SZ survey with a flux limit of $5$ mJy.}
\end{figure}
It is evident that the imposed flux-limit together with the scarcity
of high-mass halos at large redshifts, induces an abrupt decline of
the ${\cal N}(z)$ with $z$, especially in the case of the cluster
X-ray survey. As can be seen from the left panel of
\newnewnewtext{Fig.\,5}, this fact cancels the possibility of observing the
intrinsic halo number-count differences between the different models
($\Lambda$, $\Lambda_{RG}$ and $\Lambda_{n}$).
However, the expected SZ cluster
number-counts (Fig. 6) show that we maybe able to detect significant
differences in the redshift range $2.5\mincir z\mincir 3$ (at higher
redshifts only a very small number of SZ clusters will be detected)
at a level of $\sim 6-12\%$, which translates in number count
differences, over the whole sky, of $\sim 100$ clusters.

\section{Conclusions}
In this paper, we have studied analytically and numerically the
overall dynamics of several time varying  vacuum
scenarios for spatially flat FLRW geometry, beyond the usual
constant vacuum model (the
standard $\CC$CDM cosmology). We wish to spell out
clearly which are the basic assumptions and conclusions of our
analysis.

\begin{itemize}
\item We use various functional forms in order to parameterize the
vacuum energy density. In particular, we consider: (a) a vacuum
model based on the renormalization group (RG) in quantum field
theory, $\Lambda_{RG}=n_0+ n_2\,H^{2}$; (b) a vacuum model
in which $\CC$ evolves proportional to the total energy density of
the universe, hence $\CC_{H_1}\propto H^2$; (c) a power series
expansion in $H$ up to the second order without constant term,
$\Lambda_{PS_{1}}=n_{1}H+n_{2}H^{2}$; (d) a vacuum energy density
which decays proportional to the Hubble expansion rate as
$\Lambda_{PS_{2}}=n_{1}H$; and (e) a vacuum energy density which
evolves with an arbitrary power of the scale factor as
$\Lambda_{n}=3\gamma(3-n)a^{-n}$. These models have
different theoretical motivations, but not all of them are able to
withstand the stringent experimental tests provided by the various
observational sources. In this framework, we first perform a joint
likelihood analysis in order to put tight constraints on the main
cosmological parameters by using the current observational data
(SNIa, BAOs and CMB shift parameter, together with the growth rate
of galaxy clustering). Also, we find that the above models can
accommodate a late time accelerated expansion.

\item In the case of the power series models $\Lambda_{PS_{1}}$ and
$\Lambda_{PS_{2}}$, we find that the amplitude and the shape of the
linear density contrast are for both significantly
different with respect to those of
the $\Lambda$, $\Lambda_{RG}$ and $\Lambda_{n}$ models. We also find
that for $z\le 0.5$ the matter fluctuation field of the power series
models practically freezes out, which indicates that the corresponding growth
rate of clustering tends to zero. The latter result is ruled out by
the observed growth rate of clustering measured from the optical
galaxies. In the particular case of $\Lambda_{PS_{2}}$,
even the fit of the cosmological parameters turns out to be of
poor quality because it is unable to adjust simultaneously the
observational data at low and high redshift.
In contrast, the
$\Lambda$, $\Lambda_{RG}$ and $\Lambda_{n}$ models match well the
observed growth rate and they provide good quality fits
of the cosmological parameters at all redshifts. We
remark that the power law model $\Lambda_{n}$ behaves effectively
as a scalar field model with an equation of state slightly tilted in
the phantom DE regime ($w<-1$). The idea that the models with a
dynamical cosmological term $\CC$ may behave effectively as
quintessence and/or phantom DE has been described in general terms
in the literature\,\cite{SS12}.

\item The particular case of the RG
model for $n_0=0$ (or of the power series model for $n_1=0$), i.e.
the $\CC_{H_1}=n_2\,H^2$ model, is ruled out at a high significance
level because the best parameter fit value for $n_2$ is incompatible
with the necessity of growing structure formation in this model.

\item In the case of $\Lambda$,
$\Lambda_{RG}$ and $\Lambda_{n}$ vacuum scenarios, the large scale
structures (such as galaxy clusters) form later ($z\sim 2$) with
respect to those produced in the framework of the $\Lambda_{PS_{1}}$
and $\Lambda_{PS_{2}}$ models ($z\ge 4$). Therefore, in view of the
observational data, the former are much more favored as compared to
the latter.

\item The expected redshift distribution of cluster-size halos in
realistic future X-ray or SZ cluster surveys indicates that we will not be
able to distinguish the closely resembling models (constant vacuum,
quantum field and power-law vacuum) based on the evolution of
cluster abundances using the first type of survey, but there
is some limited hope with the second.
\end{itemize}

{\bf Acknowledgments.}  We are grateful to J. Grande for a careful
reading of the manuscript and for useful discussions, and also to A.
Avgoustidis for clarifying a point on model independent
determinations of the cosmological parameters. We are indebted to
F.R. Klinkhamer and G.E. Volovik for detailed correspondence on some
of the theoretical motivations for the model that we have denoted
$\Lambda_{PS_{2}}$. Last but not least, we are thankful to J. Fabris
for reading the manuscript and for useful comments. JS has been
supported in part by MEC and FEDER under project FPA2007-66665, by
the Spanish Consolider-Ingenio 2010 program CPAN CSD2007-00042 and
by DIUE/CUR Generalitat de Catalunya under project 2009SGR502. MP
acknowledges funding by Mexican CONACyT grant 2005-49878.

\appendix
\section{Alternative view of the basic cosmological equations}
The goal in this appendix is to give the reader the opportunity to
appreciate the relative similarities and differences among the main
time varying vacuum models ($\Lambda_{RG}$, $\Lambda_{PS_{1}}$ and
$\Lambda_{n}$). In particular, we re-write the basic cosmological
equations in terms of a new set of formal cosmological
parameters $(\tilde{\Omega}_{m},\tilde{\Omega}_{\Lambda})$. In this
new space, the basic structure of the $\CC$CDM Hubble flow
(\ref{hub1}) is more closely preserved, in the sense that it appears
as the sum of $\,\tilde{\Omega}_{\Lambda}$ times a function (which
can just be one in some case) and another term involving
$\tilde{\Omega}_{m}$ times another function. The two functions are
normalized to one at present, so that the sum rule
$\tilde{\Omega}_{m}+\tilde{\Omega}_{\Lambda}=1$ is fulfilled. The
change of basis from the physical parameters $(\Omm,\OL)$ to the
formal $(\tilde{\Omega}_{m},\tilde{\Omega}_{\Lambda})$ ones involves
additional parameters $\gamma_i$ of the current model. The two
parameter spaces coincide only if we can set the values of the
$\gamma_i$ such that the vacuum becomes static.  Consider the
following examples.
\begin{itemize}
\item $\Lambda_{RG}$ model: If we use eq.(\ref{otran1}),
then the basic cosmological equations (see section 4.2) take the following forms:
\begin{equation}
H(t)=\sqrt{\tilde{\Omega}_{\Lambda}}\;H_{0}
\;{\rm coth}\left[\frac{3H_{0}(1-\gamma)\sqrt{\tilde{\Omega}_{\Lambda}}}{2}\;t\right]
\end{equation}
and
\begin{equation}
a(t)=\left(\frac{\tilde{\Omega}_{m}}{\tilde{\Omega}_{\Lambda}}\right)^{\frac{1}{3(1-\gamma)}}
\sinh^{\frac{2}{3(1-\gamma)}}
\left[\frac{3H_{0}(1-\gamma)\sqrt{\tilde{\Omega}_{\Lambda} }}{2}\;t\right]
\end{equation}
or
\begin{equation}
E^{2}(a)=\tilde{\Omega}_{\Lambda}+\tilde{\Omega}_{m}a^{-3(1-\gamma)} \;\;.
\end{equation}
The scale factor at the inflection point is
\begin{equation}
a_{I}=
\left[\frac{(1-3\gamma)\tilde{\Omega}_{m}}{2\tilde{\Omega}_{\Lambda}}\right]^{1/3(1-\gamma)}\;\;
\end{equation}
Obviously, in the $(\tilde{\Omega}_{m},\tilde{\Omega}_{\Lambda})$
basis, the above equations generalize more tightly those of the
concordance $\Lambda$ cosmology (see section 4.1)
and reduce exactly to them for $\gamma=0$.

\item $\Lambda_{PS_{1}}$ model: If we use eq.(\ref{otran2}), then the corresponding
basic cosmological equations (see section 4.4) become:
\begin{equation}
H(t)=\tilde{\Omega}_{\Lambda}H_{0}\;
\frac{{\rm e}^{\gamma \tilde{\Omega}_{\Lambda} H_{0}t/2}}
{{\rm e}^{\gamma \tilde{\Omega}_{\Lambda} H_{0}t/2}-1}
\end{equation}
and
\begin{equation}
 a(t)=\left(\frac{\tilde{\Omega}_{m}}{\tilde{\Omega}_{\Lambda}}\right)^{2/\gamma}
\left({\rm e}^{\gamma \tilde{\Omega}_{\Lambda}
H_{0}t/2}-1\right)^{2/\gamma}
\end{equation}
or
\begin{equation}
E(a)=\tilde{\Omega}_{\Lambda}+\tilde{\Omega}_{m}a^{-\gamma/2}\;\;.
\end{equation}
Notice that in this particular case it is $E(a)$, rather than
$E^2(a)$, which appears decomposed as a sum of two terms in the new
parameter space. A perfect analogy with the $\CC$CDM Hubble flow is
not always possible. Indeed, in this model the vacuum can never
coincide with that of the standard model, except for the trivial
(and excluded) situation where its energy is zero. The scale factor
at the inflection point is given by
\begin{equation}
 a_{I}=\left[\frac{(\gamma-2)\tilde{\Omega}_{m}}{2 \tilde{\Omega}_{\Lambda}}\right]^{2/\gamma} \;\;.
\end{equation}

\item $\Lambda_{n}$ model: In this case we utilize eq.(\ref{otran4}). Therefore,
the normalized Hubble flow obeys
\begin{equation}
E^{2}(a)=\tilde{\Omega}_{m}a^{-3}+\tilde{\Omega}_{\Lambda}a^{-n}\,,
\end{equation}
while the corresponding inflection point is
\begin{equation}
a_{I}=
\left[\frac{\tilde{\Omega}_{m}}{(2-n)\tilde{\Omega}_{\Lambda}}\right]^{1/(3-n)}
\;\;.
\end{equation}
It is worth noting, that the current $\Lambda_{n}$ cosmological
model can be viewed as a classical {\em quintessence} model
($P_{Q}=w\rho_{Q}$, with $w=-1+n/3$), as far as the global dynamics
is concerned. Since, however, $n<0$ (i.e. $w<-1$) is
preferred by the data, in practice it behaves effectively as phantom
DE. The standard $\CC$CDM cosmology is recovered from this model in
the limit $n\to 0$ and $\gamma=\OL H_0^2/3$.

\end{itemize}

\section{The $\Lambda_{RG}$ growth factor}
With the aid of the differential equation theory we present the
growing model solution that is relevant to eq.(\ref{eq:222}) for the
RG model. Since the variable
introduced in (\ref{tran11}) satisfies $y>1$, it is possible to find
the solution $D(y)$ of the differential equation (\ref{eq:222}) in
terms of the associated Legendre functions of the second kind,
${\cal Q}_{\nu}^{\mu}(y)$. The appropriate transformation reads as
follows,
 \be \label{ap1}
D(y)=(y^{2}-1)^{\frac{5-3\beta}{6\beta}} {\cal Q}_{\nu}^{\mu}(y)\,,
\;\;\; \nu=\frac{1}{3\beta}\,, \;\;\; \mu=\nu-1 \ee where
\be \label{ap2} {\cal Q}_{\nu}^{\mu}(y)={\cal C}
y^{-\nu-\mu-1}(y^{2}-1)^{\frac{\mu}{2}} F(a,b,c,\frac{1}{y^{2}}) \ee
with $a=1+\frac{\nu}{2}+\frac{\mu}{2}$,
$b=\frac{1}{2}+\frac{\nu}{2}+\frac{\mu}{2}$, $c=\nu+\frac{3}{2}$ and
\be {\cal C}={\rm e}^{i\mu
\pi}2^{-\nu-1}\sqrt{\pi}\frac{\Gamma(\nu+\mu+1)}
{\Gamma(\nu+\frac{3}{2})} \;\;. \ee Inserting eq.(\ref{ap2}) into
eq.(\ref{ap1}) and after some algebra we have \be D(y)={\cal
C}(y^{2}-1)^{\frac{1-\beta}{\beta}} y^{-\frac{2}{3\beta}}
F\left(\frac{1}{3\beta}+\frac{1}{2},\frac{1}{3\beta},\frac{1}{3\beta}+\frac{3}{2},
\frac{1}{y^{2}}\right) \ee or
\begin{equation}
D(y)={\cal C}(y^{2}-1)^{\frac{4-9\beta}{6\beta}} y
F\left(\frac{1}{3\beta}+\frac{1}{2},\frac{3}{2},
\frac{1}{3\beta}+\frac{3}{2},-\frac{1}{y^{2}-1} \right)
\end{equation}
where we have used the well known linear transformation formula:
\be
F(\alpha,b,c,x)=(1-x)^{-\alpha}F\left(a,c-b,c,\frac{x}{x-1}\right) \;\;.
\ee
Note, that in our formulation the parameters become:
$\alpha=\frac{1}{3\beta}+\frac{1}{2}$, $b=\frac{1}{3\beta}$,
$c=\frac{1}{3\beta}+\frac{3}{2}$ and $x=\frac{1}{y^{2}}$.

\vspace{0.5cm}
\section{Unification of the vacuum models}
In this appendix we examine a more general class of vacuum models:
\begin{equation}\label{unified}
\Lambda=n_{0}+n_{1}H+n_{2}H^{2}\,,\ \ (n_{0}\ge 0)\,.
\end{equation}
The time evolution equation for the Hubble flow is obtained by
eq.(\ref{frie34}) as: \be \label{eq:799}
\int_{+\infty}^{H}\frac{dy}{-\lambda y^{2}+n_{1} y+n_{0}}
=\frac{t}{2} \ee where $\lambda=3-n_{2}$. If we assume that $\lambda
> 0$ (or $n_{2}<3$) then corresponding general solution of
eq.(\ref{eq:799}) is \be \label{eq:899} H(t)=\frac{\rho_{2}\;{\rm
e}^{\lambda(\rho_{2}-\rho_{1})t/2}-\rho_{1}} {{\rm
e}^{\lambda(\rho_{2}-\rho_{1})t/2}-1} \ee and \be \label{eq:999}
a(t)=a_{1}\left({\rm e}^{\lambda(\rho_{2}-\rho_{1})t/2}-1
  \right)^{2/\lambda}
{\rm e}^{\rho_{1}t}
\ee
where
\be
\rho_{2,1}=\frac{n_{1}\pm \sqrt{D}}{2\lambda} \;\;\; \rho_{2}> \rho_{1}
\ee
and $D=n^{2}_{1}+4\lambda n_{0} \ge 0$ is
the discriminant.

Obviously, the $\Lambda_{PS_{1,2}}$,
$\Lambda_{RG}$ and $\Lambda$ models
are particular solutions of the general vacuum model.
Indeed, we have:

\begin{itemize}
\item {\it Case 1:}
If $n_{0}=0$ then $\rho_{1}=0$ and $\rho_{2}=n_{1}/\lambda$. In this
case, the basic cosmological equations [see eq.(\ref{eq:899}) and
eq.(\ref{eq:999})] reduce to those found by either the
$\Lambda_{PS_{1}}$ (for $\lambda=3-n_{2}>0$) or the
$\Lambda_{PS_{2}}$ (for $\lambda=3$) model respectively (see
eq.\,\ref{frie4}).

\item {\it Case 2:}
If $n_{1}=0$ then $\rho_{2}=-\rho_{1}$.
We select the unknown constants such as
$n_{0}=3\Omega_{\Lambda}H^{2}_{0}(1-\gamma)$ and
$n_{2}=3\gamma$.
Therefore, the general Hubble expansion eq.(\ref{eq:899}),
reduces to that derived by the
$\Lambda_{RG}$ model (see eq.\ref{frie455}).

\item {\it Case 3:}
If $(n_{1},n_{2})=(0,0)$ then $\rho_{2}=-\rho_{1}=\sqrt{3n_{0}}/3$.
Thus, for $n_{0}=3\Omega_{\Lambda}H^{2}_{0}$, the Hubble expansion eq.(\ref{eq:899}),
reduces to that derived by the
concordance $\Lambda$ cosmology (see eq.\ref{frie556}).

\end{itemize}
The analysis of structure formation for the model (\ref{unified}) in the most general case when all
the coefficients $n_i$ are non-vanishing cannot be performed
analytically. We shall report on this case elsewhere\,\cite{BGPS}.


\end{document}